\documentclass[preprint]{aastex}
\usepackage{emulateapj5,epsfig}

\newcommand{\te}{\ensuremath{\mathrm{T}_{eff}}}
\newcommand{\ebv}{{\ensuremath{E(\mathrm{B}-\mathrm{V})}}}
\newcommand{\wfpc}{\emph{WFPC2}}
\newcommand{\ha}{\ensuremath{\mathrm{H}\alpha}}
\newcommand{\hst}{\emph{HST}}
\newcommand{\ie}{\emph{i.e.}\ }
\newcommand{\eg}{\emph{e.g.,}\ }
\newcommand{\pp}{\ensuremath{^{\prime\prime}}}
\newcommand{\mg}{\emph{mag}}

\newcommand{\eij}[2]{{\ensuremath{E(\mathrm{#1}-\mathrm{#2})}}}
\newcommand{\bi}[1]{{\ensuremath{\mathrm{#1}}}}
\newcommand{\q}{\ensuremath{Q_\mathrm{UBI}}}
\newcommand{\ra}{\ensuremath{\rightarrow}}
\newcommand{\dfive}{\ensuremath{\overline{\delta}_5}}
\newcommand{\efig}[3]{\centerline{\epsfig{file=#1, height=#2, angle=#3}}}
\newcommand{\dfig}[6]{\centerline{
\epsfig{file=#1, width=#2\linewidth, angle=#3}
\epsfig{file=#4, width=#5\linewidth, angle=#6}
}}

\slugcomment{Accepted for publication in the February 2002 issue of AJ}

\shorttitle{Accurate stellar population studies}
\shortauthors{Romaniello et al}

\begin{document}

\title{Accurate Stellar Population Studies \\from Multiband Photometric
Observations$^\dag$}
\renewcommand{\thefootnote}{\fnsymbol{footnote}}\footnotetext[2]{Based
on observations with the NASA/ESA Hubble Space Telescope, obtained at
the Space Telescope Science Institute, which is operated by
AURA,~Inc., under NASA contract
NAS~5-26555.}\renewcommand{\thefootnote}{\arabic{footnote}}

\author{Martino Romaniello}
\affil{ESO, Karl-Schwarzschild-Stra{\ss}e 2, D-85748 Garching bei
M\"unchen, Germany} 
\email{mromanie@eso.org}
\author{Nino Panagia\altaffilmark{1} }
\affil{Space Telescope Science Institute, 3700 San Martin Drive,
Baltimore, MD~21218} 
\email{panagia@stsci.edu}
\author{Salvatore Scuderi}
\affil{Osservatorio Astrofisico di Catania, Viale A. Doria 6, I-95125,
Catania, Italy}
\email{scuderi@sunct.ct.astro.it}
\and
\author{Robert P. Kirshner}
\affil{Harvard-Smithsonian Center for Astrophysics, 60 Garden Street,
Cambridge, MA~02138} 
\email{kirshner@cfa.harvard.edu}
\altaffiltext{1}{On assignment from the Research and Scientific Support
Department of ESA}

\begin{abstract}
We present a new technique based on multi-band near ultraviolet and optical
photometry to measure both the stellar intrinsic properties, \ie
luminosity and effective temperature, and the interstellar dust
extinction along the line of sight to hundreds of stars per square
arcminute. The yield is twofold. On the one hand,  the resulting
reddening map has a very high spatial resolution, of the order of a few
arcseconds, and can be quite effectively used in regions where the
interstellar material is patchy, thus producing considerable
differential extinction on small angular scales. On the other hand,
combining the photometric information over a wide baseline in
wavelength provides an  accurate determination of temperature and
luminosity for thousands of stars. As a test case, we present the
results for the region around Supernova~1987A in the Large Magellanic
Cloud imaged with the \wfpc\ on board the Hubble Space Telescope.
\end{abstract}


\section{Introduction}
Stars provide most of the light we detect in the Universe and the entire
chemical and dynamical evolution of galaxies is strongly influenced by the
formation of many generations of stars. The most massive ones enrich the
surrounding medium of highly processed material and supply vast amounts of
kinetic energy (both with their winds and, even more, with their ultimate
explosions as Type~II Supernov\ae). Lower mass stars constitute one of
the main sources of crucial elements, such as Carbon and Nitrogen
(through stellar winds first, and planetary nebula ejection later) in
their individual evolution, and may be the main suppliers of the iron
group elements if, in binary systems, they succeed to merge and explode
as Type~Ia Supernov\ae.

A major problem that one encounters almost ubiquitously when studying
stars is the effect that interstellar dust has on the  observed
spectral distribution of their light.   A precise measurement of the
interstellar reddening along the line of sight to the stars under study
is the first step, and a crucial one, towards determining the intrinsic
properties of the stellar populations. All sky reddening maps, such as
the one derived from dust temperature by \citet{sch98}, for example,
may not be satisfactory for detailed stellar populations studies. For
example, it cannot be used in the direction of nearby galaxies, such
as the Magellanic Clouds and M31. Also, the derived value of the dust
column density is extremely sensitive to dust temperature and, in any
case, not accurate for high values of extinction
\citep[$A_V>0.5$~\mg,][]{arce99}. Moreover, the  angular resolution
of the map is pretty coarse ($\sim6^{\prime}$, as compared to a few
arcseconds for the method presented here) and, thus, it does not allow
to cope effectively with the patchy structure typical of the
interstellar medium. 

It is certainly preferable to determine the amount of dust in front of
the stars from the light coming from the stars themselves. While
spectroscopic data may certainly provide an accurate determination of
the reddening and the stellar parameters, the multiplexing
capabilities of current spectrographs are limited, at best, to a few
tens of (relatively bright) objects per telescope pointing. The method
we describe in this paper overcomes this limitation by using only
imaging and, thus, providing measurements of both the stellar
parameters and the reddening along the line of sight to thousands of
stars per pointing. The gain in terms of statistics is obviously
tremendous. As a consequence, the spatial resolution of the resulting
extinction map is extremely high, with an \ebv\ measurement every few
arcseconds, even for moderately dense star fields. While here we test
this newly developed technique on \hst-\wfpc\ observations, the method
can be applied to data obtained with any instrument. The quality of the final
result, however, will obviously depend on the quality of the input photometry.

The whole purpose of this paper is to describe a method to retrieve the
stellar parameters from this kind of data by fitting model atmospheres to
the observed fluxes.
In section~\ref{sec:data-red} we describe the dataset considered
and  the reduction techniques adopted. In section~\ref{sec:cmd} the
limitations of the classical approach based on Color-Magnitude diagrams
are discussed and the method we have developed to overcome them via a
multiband fit to the observed spectra is presented in section~\ref{sec:proc}.
The results of this procedure are discussed in section~\ref{sec:fresults},
while section~\ref{sec:meth_gte} deals with the effects that the gravity and
metallicity of the input model spectra have on the derived parameters.
Finally, the spectral coverage needed to recover bot \te\ and \ebv\ is
explored in section~\ref{sec:need} and the role of random and systematic
errors is assessed in section~\ref{sec:err}.

The broad band colors in selected \wfpc\ filters and the corresponding
extinction coefficients, which are needed for any application of our
procedure, are provided in appendix~\ref{sec:app_colors}.

\section{Observations and Data Reduction}\label{sec:data-red}
Our method was developed and tested using \hst-\wfpc\ multiband
observations. In particular, we have taken advantage of the wealth of
information provided by the repeated observations of Supernova~1987A (SN1987A)
and the field surrounding it in the Large Magellanic Cloud (LMC). Since
1994, Supernova~1987A was imaged with the \wfpc\ every year as a part
of the {\bf S}upernova {\bf IN}tensive {\bf S}tudy project (SINS, PI:
Robert P. Kirshner). The log of the observations we have used is
reported in Table~\ref{tab:log}. The \wfpc\ filters do not exactly match any
of the ground-based photometric systems and the closest analog of the
passbands we have used is reported in the last column of this Table.
The observations cover an almost circular portion of the sky
centered on SN1987A and with a radius of $130\pp$, corresponding to
roughly 32~pc at the distance of the LMC \citep[51.4~kpc,] []{pan99, 
rom00}. The true color image derived from our data is shown in
Figure~\ref{fig:truecolor}.

\placetable{tab:log} 
\placefigure{fig:truecolor}

The observations were processed through the standard PODPS (Post
Observation Data Processing System) pipeline for bias removal and flat
fielding. In all the cases the available images for each filter were
combined to remove cosmic rays events.  The long baseline in wavelength
that they cover (more than a factor of 4, extending from $\sim$2,300 to
$\sim$9,600~\AA) together with the unprecedented spatial resolution of
\wfpc\ ($\sim0\pp.1$, \ie roughly 0.03~pc at the distance of the LMC) make
of this dataset a unique and ideal one to investigate the problem of
disentangling the effects of reddening from those of different
effective temperatures for stars of different spectral type.

The plate scale of the camera is 0.045 and 0.099 arcsec/pixel in the PC
and in the three WF chips, respectively. We performed aperture
photometry following the prescriptions by \citet{gil90} as refined by
\citet{rom98}, \ie measuring the flux in a circular aperture of
2~pixels radius and the sky background value in an annulus of internal
radius 3~pixels and width 2~pixels. Aperture photometry is perfectly
adequate for our study because crowding is never a problem in our
images. Actually, the average separation of stars from each other is
about 1.3\pp\ (\ie $\sim29$ pixels in the PC chip and $\sim13$ pixels
in the WF ones) which is much larger than the \wfpc\ PSF width. We have
used the \emph{IRAF-synphot} synthetic photometry package to compute
the theoretical Data Numbers to be compared to the observed ones. The
\wfpc\ calibration and, consequently, the \emph{synphot} results are
typically accurate to within $\pm$ 5\% \citep{whi95,bagg97}.

We identify the stars in the deepest frame (F555W, \ie V band-like
filter) and then measure their magnitudes in all of the other filters.
The total number of stars identified in this way is 21,955. If a star is
not detected in a band, then its magnitude in that band is substituted by
a $3\sigma$ upper limit. For about 12,340 of them, \ie 56\%, the photometric
accuracy is better than 0.1~\mg\ in the F555W, F675W and F814W filters. The
number of stars with accuracy better than 0.1~\mg\ drops to 6,825 (31\%) in
the F439W band, and only 786 (3.6\%) stars have  an uncertainty in the F255W
filter smaller than 0.2~\mg.

The brightest stars in each CCD chip are saturated (\ie most stars
brighter than 17.5 in the F555W passband). We recover their photometry
either by fitting the unsaturated wings of the PSF for moderately
saturated stars, \ie  with no saturation outside the central 2 pixel
radius, or by following the method developed by \citet{gilli94} for
heavily saturated stars. In most cases, the achieved photometric
accuracy is better than 0.05 magnitudes. Full description of the
methods used can be found in \citet{rom98}. As a sanity check we have
compared our photometry in the F439W and F555W bands with ground-based
one in B and V \citep[28 stars brighter than $V\simeq20$
within 30\pp\ from SN1987A,][]{ws90} and found an overall excellent
agreement, \ie {\it rms} deviations less than 0.05~\mg\ in both filters,
except for the obvious cases of few stars that appear as point-like in
ground based images but are resolved into more than one object on the
\wfpc\ frames. This result confirms the accuracy of our photometric
measurements and the quality of the \wfpc\ calibration.

We shall see in section~\ref{sec:proc} that a good photometric
quality in different bands is needed in order to recover \ebv\ and
stellar temperature. In order to select stars with a good overall
photometry we define the mean error in 5 bands (\dfive)
(\ie excluding the UV F255W band) as:

\begin{equation}
\dfive=\sqrt\frac{\delta_{\mathrm{F336W}}^2+
\delta_{\mathrm{F439W}}^2+\delta_{\mathrm{F555W}}^2+
\delta_{\mathrm{F675W}}^2+\delta_{\mathrm{F814W}}^2}{5}
\label{eq:mer}
\end{equation}

In addition to the main field described above, we have also used a
second set of \wfpc\ images pointing $8^\prime$ south-east of SN1987A.
As shown in Table~\ref{tab:sn87a_parf_log} in this case we only use 4
filters and the F336W and F439W passbands are substituted with their
broader counterparts F300W and F450W, respectively. Being broader, the
latter are more efficient and allowed us to save precious telescope
time, always a crucial issue, without losing any vital information (see
section~\ref{sec:need}).

\placetable{tab:sn87a_parf_log}

\section{Color-Magnitude diagrams}\label{sec:cmd}
As it can be seen in Figure~\ref{fig:truecolor} Supernova~1987A and its
rings are right at the center of the field imaged with \wfpc\ and belong to
a loose cluster of blue stars \citep[see][]{prsk00}. Also, the interstellar
medium, as traced by the \ha\ emission coded in red in
Figure~\ref{fig:truecolor}, is far from being spatially uniform and its
distribution varies on scales of few arcseconds (let us recall here
that 1\pp\ corresponds to roughly 0.25~\emph{pc} at the distance of the
LMC). Since the spatial distribution of dust usually follows the one
of gas \citep[see, for
example,][]{mat90}, one can expect the interstellar reddening to vary
on similar, if not even smaller, scales. This is further confirmed by
an inspection of Figure~\ref{fig:cmd} where we show the Color-Magnitude
diagrams (CMD) for different combination of bands.

\placefigure{fig:cmd}
\placefigure{fig:phot_err}

The black dots in each CMD in Figure~\ref{fig:cmd} are the 6,695 stars
with $\dfive<0.1$~\mg, where \dfive\ is the mean error defined in
equation~(\ref{eq:mer}). The error threshold of $\dfive<0.1$ reflects
itself as a magnitude threshold at
\mbox{$m_{\mathrm{F555W}}\simeq~23$}. We estimate that the completeness at
this magnitude limit is very close to 100\% because the density of
detected stars down to 23th magnitude is rather low, \ie about 1 star
per 4.6 square-arcsec area \citep[][]{prsk00}.

It is apparent that, despite the high quality of the measurements
(internal uncertainties less than 0.1 magnitudes, see also
Figure~\ref{fig:phot_err}), the various features
of the CMDs, such as the Zero Age Main Sequence for the early type
stars and the Red Giant clump for the more evolved populations, are
rather ``fuzzy'' and not sharply defined. Although this is due in part
to the presence of several stellar populations projected on each other
\citep[][]{prsk00}, most of the problem arises from the fact that
reddening is not quite uniform over the field, thus causing the points
of otherwise identical stars to fall in appreciably different locations
of the CMDs.

\section{The fitting procedure}\label{sec:proc}
\begin{figure}
\efig{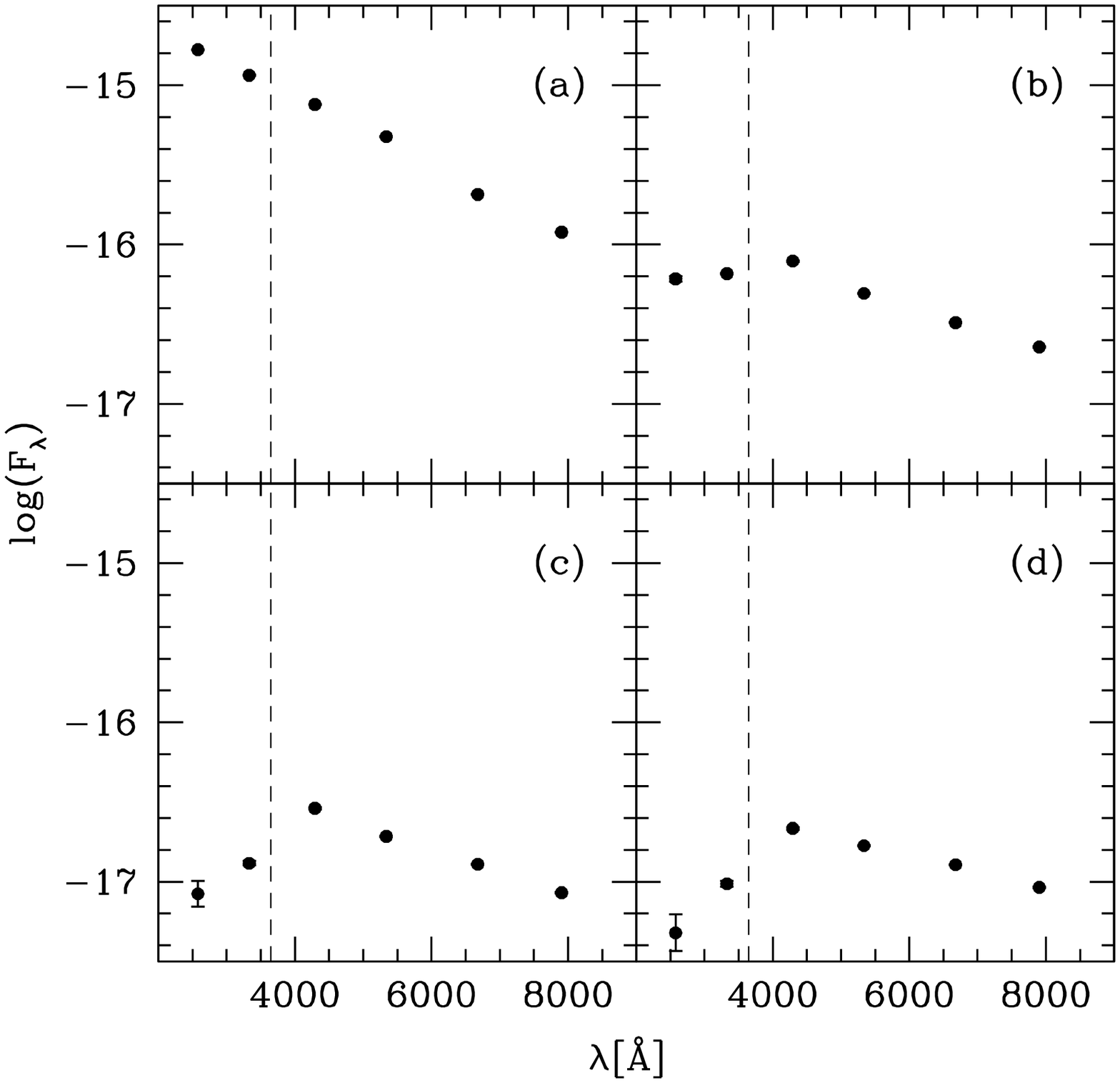}{0.9\linewidth}{0}
\caption{Four examples of the spectral resolution of the
  ``Wide Band spectroscopy''. The magnitudes and errors of these four stars
  are given in Table~\protect{\ref{tab:wbsp_nfit}}. The
  position of the Balmer jump at 3646~\AA\ is indicated with a dashed
  line.\label{fig:wbsp_nfit}}
\end{figure}

The large number of bands available in our dataset provides a sort of
\emph{wide-band spectroscopy} which defines the continuum spectral
distribution of each star quite well. Of course, the resolution is very
low, of the order of $\lambda/\Delta\lambda\simeq~7$, but, still,
important features, such as the Balmer jump between the  \hst\  F336W and 
F439W bands, are clearly visible. Four examples of ``wide band spectra''
are shown in Figure~\ref{fig:wbsp_nfit} (the flux F$_\lambda$ is expressed
in units of erg~cm$^{-2}$~s$^{-1}$~\AA$^{-1}$) and their observed
magnitudes are listed in Table~\ref{tab:wbsp_nfit}. The whole purpose of
this paper is to describe a method to retrieve the stellar parameters
from this kind of data by fitting model atmospheres to the observed
fluxes.

\placefigure{fig:wbsp_nfit}
\placetable{tab:wbsp_nfit}

The \emph{shape} of the observed spectrum is  mostly  determined by
the effective temperature of the star and by the amount of dust along
the line of sight. The other quantity of interest, namely  the solid
angle subtended by the star (or, correspondingly, its angular radius) is
obtained by a straight  normalization of the spectral distribution. 

The general idea of our method is to recover simultaneously the
extinction along the line of sight and the effective temperature of a
star by fitting the observed magnitudes to the ones computed using stellar
atmosphere models reddened by various amounts of \ebv. The
angular radius, then, is directly obtained from the ratio of the observed
flux, dereddened with the best fitting \ebv\ value, to the one at the
surface of the star, as given by the best fitting model atmosphere.
Thus, the two fundamental ingredients that go into the fit are the
\emph{stellar  atmosphere models} and the \emph{reddening law}: we have
adopted the models by \citet{bes98} and the reddening law by
\citet{scud96}.

We have chosen the models by \citet{bes98}, as they provide a well
tested, homogeneous set covering the temperature interval between 3,500
and 50,000~K in the wavelength range between 90~\AA~ and 160~$\mu m$.
They are computed with an updated version of the \texttt{ATLAS9} code
used for the \citet{kur93} models with no allowance for convective
overshooting, as discussed by \citet{fcast97a,fcast97b}.

We have used the IRAF task \emph{synphot} to compute the theoretical
magnitudes at the surface of the star for the grid of temperatures
provided by \citet{bes98} by convolving the model spectra with the
\hst+\wfpc\ response curves. Once the convolution is performed, we use
a spline interpolation to get a temperature grid with a logarithmic
step of 0.005, \ie 1\%.  We have used model atmospheres with
$\log(g)=4.5$ (cgs units), which is the typical value for Main Sequence
stars, and Z$=0.3\cdot$Z$_\sun$, the mean expected metallicity of the
LMC. In section~\ref{sec:meth_gte} we will  analyze and discuss the
effects of these two parameters on the fit and we  will see that
neither of them affects the values of the derived parameters
appreciably.

In order to faithfully model the data, we have computed separately the
expected counts that a given spectrum produces on each of the four
\wfpc\ chip detectors. Also, the throughput variation between
decontaminations of the camera \citep{bagg96} was taken into account by
specifying the Julian date of the observation (the \emph{cont} keyword in
\emph{synphot}). Details of this procedure can be found in \citet{rom98},
while model magnitudes for solar and one third solar metallicity are listed
here in Tables~(\ref{tab:cas22}) and (\ref{tab:cas63}), respectively.

The other input ingredient is the \emph{reddening law}. This is a
combination of foreground extinction due to dust in the Milky Way and
internal extinction in the LMC. We have taken the first component to be
the same for all of the stars, with $\ebv_\mathrm{MW}=0.05$ \citep[see,
for example,][]{schw91} and we have used the Galactic reddening law as
compiled by \citet{savmat79}. On the other hand, the amount of
extinction from the LMC dust is one of the parameters of the fit. We
have taken the LMC reddening law in the direction of our field to be
the one determined by \citet{scud96} from a study of an \hst-FOS
ultraviolet and optical spectrum of Star~2, one of the companions to
SN1987A projected just 3\pp.9 NW of it. It is, of course,  crucially 
important to use a reddening law derived in the same field under study,
as it  may vary considerably from place to place on quite small angular
scales \citep[see, for example,][]{whit92}. These reddening laws
were used to compute the extinction coefficients in the \wfpc\
passbands for every combination of input model spectrum and value of
\ebv\ (see Table~\ref{tab:abs}).

The fitting procedure consists of two steps:

\begin{itemize}
  \item we first use a reddening-free color to determine
     which stars can be used to measure the reddening unambiguously;
  \item once this is done, we use the full information from the
     6~bands to fit the observed spectral energy distribution for all
     of the stars.
\end{itemize}

We will now describe these two steps.

\subsection{First step: reddening-free colors}\label{sec:redfree}
\begin{figure}
\efig{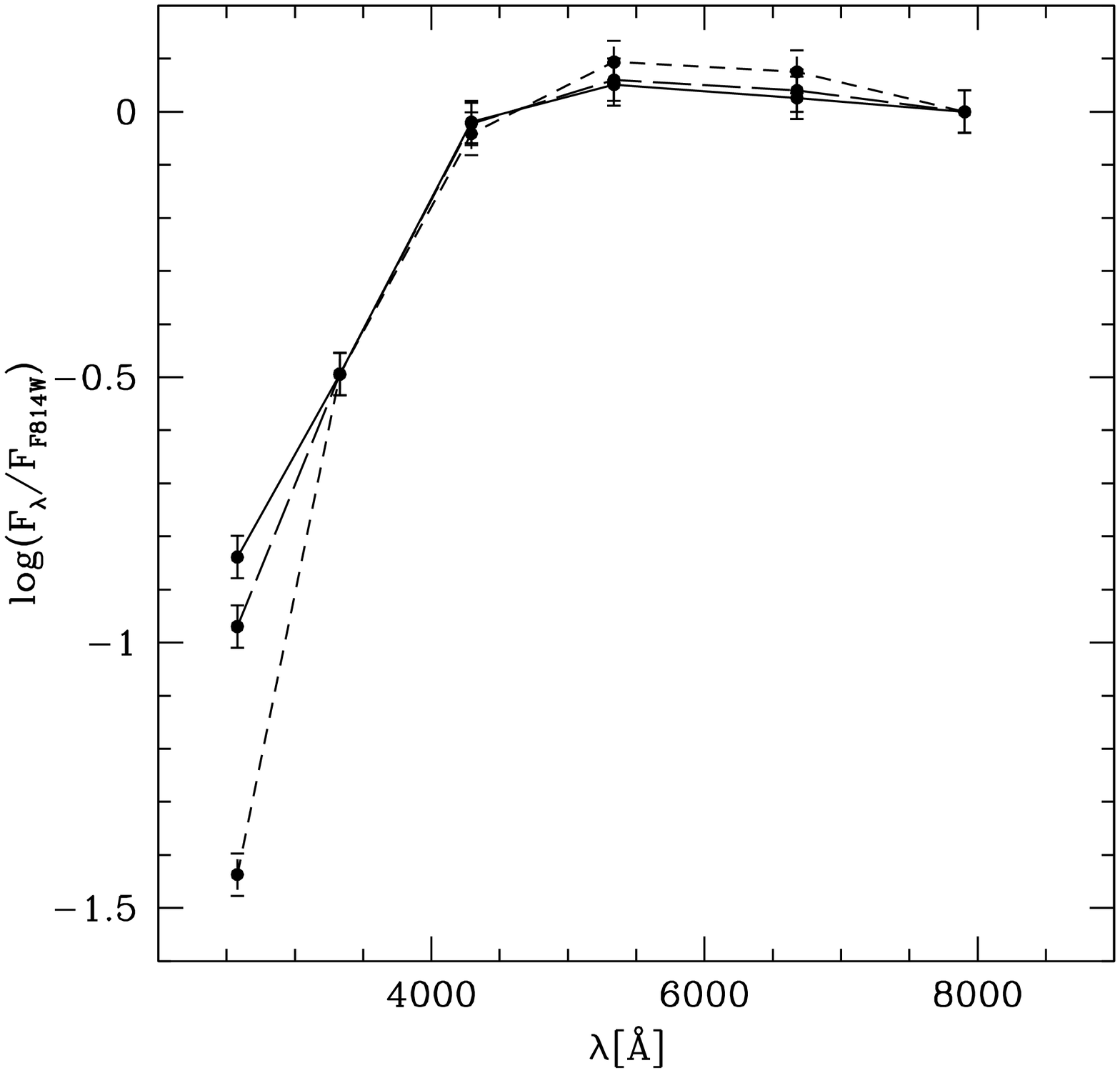}{0.9\linewidth}{0}
\caption{Comparison of spectra for different
  combinations of \te\ and \ebv: $\te=5,000$ and $\ebv=0$ (short-dash line),
  $\te=7,600$ and $\ebv=0.55$ (long-dash line), $\te=12,000$ and $\ebv=0.86$
  (full line). The spectra are normalized to the F814W flux and errorbars
  corresponding to 0.1~\mg\ are shown. As it can be seen, at this resolution
  the spectra are almost indistinguishable for wavelengths longer than
  3,300~\AA\ and ultraviolet imaging is needed to tell them apart.
  \label{fig:etdeg}}
\end{figure}

It is a well known fact that the solution in the \ebv-\te\ plane may not be
unique for stars later than, roughly speaking, A0 and that different
combinations of temperature and reddening may lead to almost indistinguishable
optical and near-UV low resolution spectra. An example is shown in
Figure~\ref{fig:etdeg} where
we compare \citet{bes98} models for three combinations of temperature and
reddening: $\te=5,000$~K and $\ebv=0$ (short-dash line), $\te=7,600$~K reddened
by $\ebv=0.55$ (long-dash line) and, finally, $\te=12,000$~K and $\ebv=0.86$
(full line). The spectra are normalized to the F814W flux and
errorbars corresponding to 0.1~\mg\ are shown. To all practical purposes the
spectra are identical for wavelengths longer than 3,300~\AA\ and only good data
in the ultraviolet can help to distinguish them. Unfortunately, though,
these data are often not available, as imaging in the ultraviolet can only be
done from space and it requires very long exposure times for red stars
\footnote{For example, 4,700 seconds would be required to reach a signal to
noise ratio of 5 in the F255W filter for an F0V star at the distance of the
Large Magellanic Cloud.}. The purpose of first step of the procedure, then,
is to use a reddening-free color to eliminate the \ebv-\te\ degeneracy when
no ultraviolet observations are available.

\placefigure{fig:etdeg}

Let us consider the reddening-free color \q, combination of the magnitudes
in the U, B and I passbands \citep[\eg][page 186]{mb81}:

\begin{equation}
\q\equiv\left(\bi{U}-\bi{B}\right)-\frac{\eij{U}{B}}{\eij{B}{I}}
\cdot\left(\bi{B}-\bi{I}\right)
\label{eq:meth_qubi}
\end{equation}
\q\ is reddening-free in the sense that:

\begin{eqnarray}
\q=\left(\bi{U}-\bi{B}\right)_0+E\left(\bi{U}-\bi{B}\right)-
\frac{\eij{U}{B}}{\eij{B}{I}}\cdot\nonumber\\
\left[\left(\bi{B}-\bi{I}\right)_0+E\left(\bi{B}-\bi{I}\right)\right] \simeq
Q_{\mathrm{UBI},0}
\end{eqnarray}
where the subscript ``0'' indicates the unreddened quantities.
\q\ is not exactly equal to $Q_{\mathrm{UBI},0}$ because the coefficient in
equation~(\ref{eq:meth_qubi}) is not a constant, but depends on the
temperature of the star and on the amount of reddening in front of it (see
Appendix~\ref{sec:app_colors}). With the reddening law of \citet{scud96}, and
using the \wfpc\ filters F336W (U), F439W (B) and F814W (I), one finds
$\eij{F336W}{F439W}/\eij{F439W}{F814W}\simeq0.42$ for $\te\gtrsim5,500$
and this is the value will use.

\begin{figure}
\efig{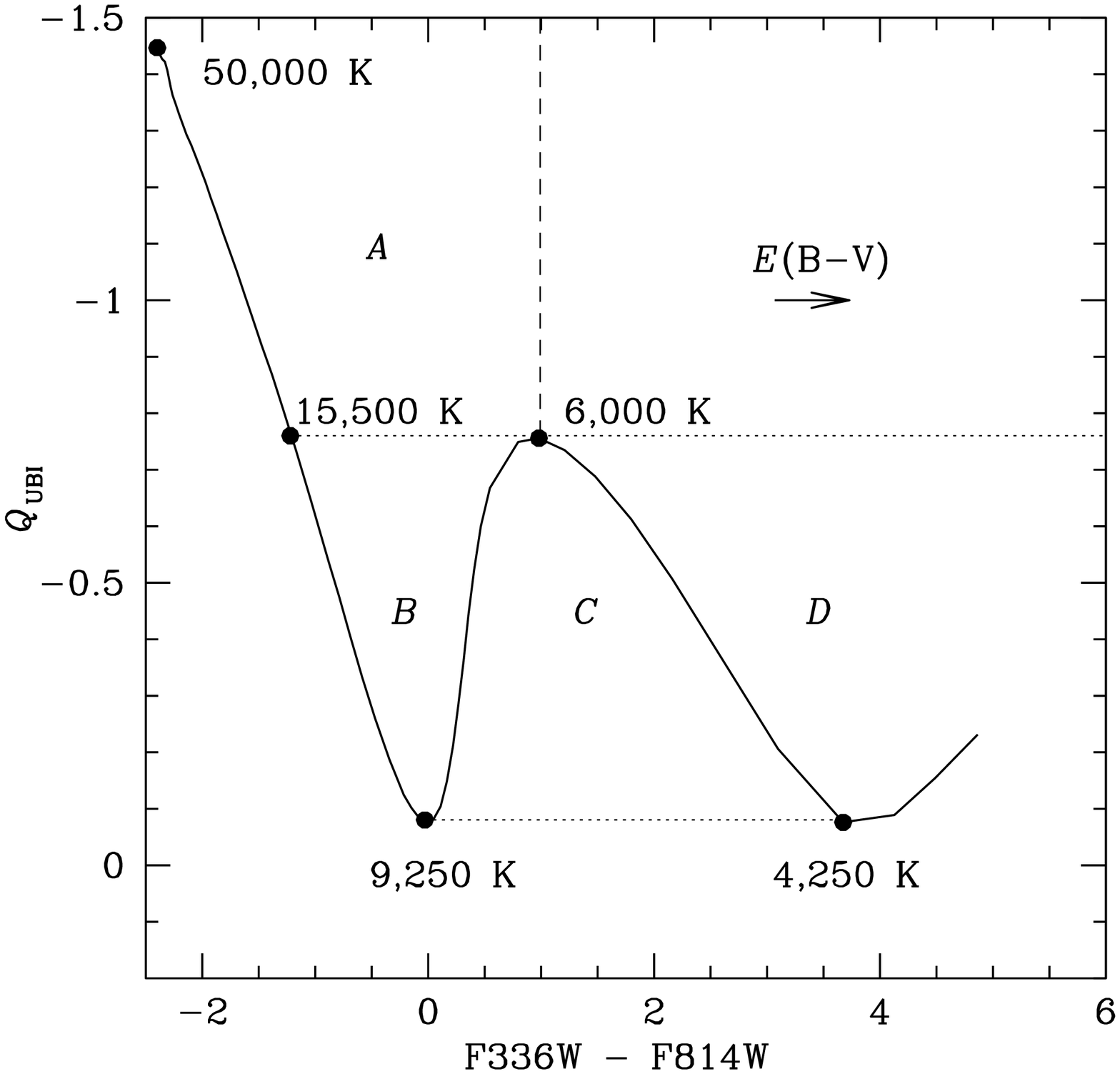}{0.9\linewidth}{0}
\caption{Reddening-free color \q\ as a function of
  $(\bi{F336W}-\bi{F814W})$ from the theoretical models by
  \protect{\citet{bes98}} for
  Z$=0.3\cdot$Z$_\sun$ and $\log(g)=4.5$. The horizontal arrow indicates the
  reddening vector for $\ebv=0.2$~\mg\ and a few important temperatures along
  the sequence are also marked. Stars in the four regions $A\ra D$ have a
  different number of solutions (see text). The dashed line highlights the
  location in this diagram of the T~Tauri stars: their colors are not well
  represented by normal photospheres.\label{fig:qubit}}
\end{figure}

The reddening-free color \q\ computed with the models by \citet{bes98}
for the WFPC2 filters is plotted in Figure~\ref{fig:qubit} versus
$(\bi{F336W}-\bi{F814W})$. The curve is for $\ebv=0$ and the arrow is the
reddening vector for $\ebv=0.2$~\mg.

\placefigure{fig:qubit}

Since, by construction, the value of \q\ does not change with \ebv, the
effect of reddening in the \q~vs~$(\bi{F336W}-\bi{F814W})$ plane is to move the
points \emph{horizontally} to the \emph{right} from the zero-reddening locus,
as indicated by the arrow in Figure~\ref{fig:qubit}. Moreover,
$\Delta(\bi{F336W}-\bi{F814W})$, the horizontal distance of an observed point
from the theoretical zero-reddening curve, is proportional to \ebv:

\begin{eqnarray}
\Delta\left(\bi{F336W}-\bi{F814W}\right)=\eij{F336W}{F814W}=\nonumber\\
\simeq 3.3\cdot\ebv
\label{eq:meth_dui}
\end{eqnarray}
whereas the observed \q\ is, to first approximation, a function
only of the star's temperature (see Figure~\ref{fig:qubit}).

As apparent from Figure~\ref{fig:qubit} \q\ is not a monotonic function
of F336W$-$F814W, \ie temperature, but it is W-shaped\footnote{The
additional hook at F336W$-$F814W$\gtrsim 3.5$ is caused by the \wfpc2\
F336W filter red leak.}. This, in turn, is due to the non monotonicity
of the Balmer jump, the intensity of which peaks for A0 stars and
declines both for earlier and later spectral types \citep[see, for
example,][]{allen73}. So, stars in region \emph{A} on
Figure~\ref{fig:qubit} have only one possible intersection with
the zero reddening curve and for them the solution \te-\ebv\ is unique.
In the case of stars in region $B$, there are three intersections with
the reddening curve, but the two on the right are easily rejected by
noticing that, according to equation~(\ref{eq:meth_dui}), they correspond to
negative values of \ebv\ and, hence, are not acceptable. The solution,
therefore, even if not unique, is unambiguous.

Things are more complicated for the stars with F336W$-$F814W$\gtrsim0$,
\ie $\te\lesssim 9,000$~K. Here, the solution is not only non-unique,
but also ambiguous. As a matter of fact, there are two (region $C$) or
three (region $D$) combinations of \te\ and \ebv\ that are absolutely
equivalent. All of these solutions are physical and there is no a
priori criterion the decide which one is the right one.

{\bf The first step of the procedure is to divide the stars in four
classes, according to their photometric error and the number of solutions in
the {\boldmath${\mathrm{T}_{eff}-E(\mathrm{B}-\mathrm{V})}$} plane.} In other
words, they are classified according to the mean error in five bands \dfive\ 
introduced in equation~(\ref{eq:mer}) and to the number of intersections that
they have \emph{to their left}, \ie bluer F336W$-$F814W color, with the
theoretical zero-reddening curve in the \q~vs.~F336W$-$F814W plane.
In order to account for the photometric errors, we project the measured value
plus or minus its error. In the following, a star belongs to a certain class
if the measured point \emph{and} the~$\pm 2\cdot\sigma$  values have the
\emph{same} number of intersections.

The classes are defined as:
  \begin{description}
    \item[class I:] stars that have only one intersection to their left. They
      are the stars in region $A$ and $B$ in Figure~\ref{fig:qubit};
    \item[class II:] stars in region $C$ in Figure~\ref{fig:qubit}. They
      have two intersections to their left. The solution for \ebv\ and \te\ is
      not unique and we shall see shortly how we choose between the different
      possibilities;
    \item[class III:] stars that, in spite of their good photometry
      ($\dfive<0.1$), have a ``problematic''
      location in the \q~vs.~F336W$-$F814W plane. These are:
      \begin{itemize}
      \item stars whose horizontal projection never intersects the
        zero-reddening curve to their left;
      \item stars that belong to region $D$ Figure~\ref{fig:qubit} and,
        hence, have three intersections to their left. As we shall discuss in
        section~\ref{sec:meth_gte}, in this region the particular combination
        of colors we have chosen is sensitive to metallicity and this
        classification scheme is not reliable;
      \item stars that fall above the peak at F336W$-$F814W$=1$, roughly along
        the long-dash line in Figure~\ref{fig:qubit}, for which the F336W flux
        is too strong as compared to the one in the other bands.
	As discussed in \citet{prsk00}, they are UV excess objects,
	most likely T~Tauri stars and the U-band
	excess is due to the presence of a circumstellar disk accreting
	onto the star. As a consequence, their spectrum is not well modeled
	by the one of a normal photosphere and no reliable \ebv\ can be
	derived by comparing their colors to those of a normal photosphere.
      \end{itemize}
      \item[class IV:] finally, stars with poor overall photometry, \ie
        $\dfive>0.1$, are assigned to this class, independent of the number
        of intersections.
  \end{description}

\begin{figure}
\efig{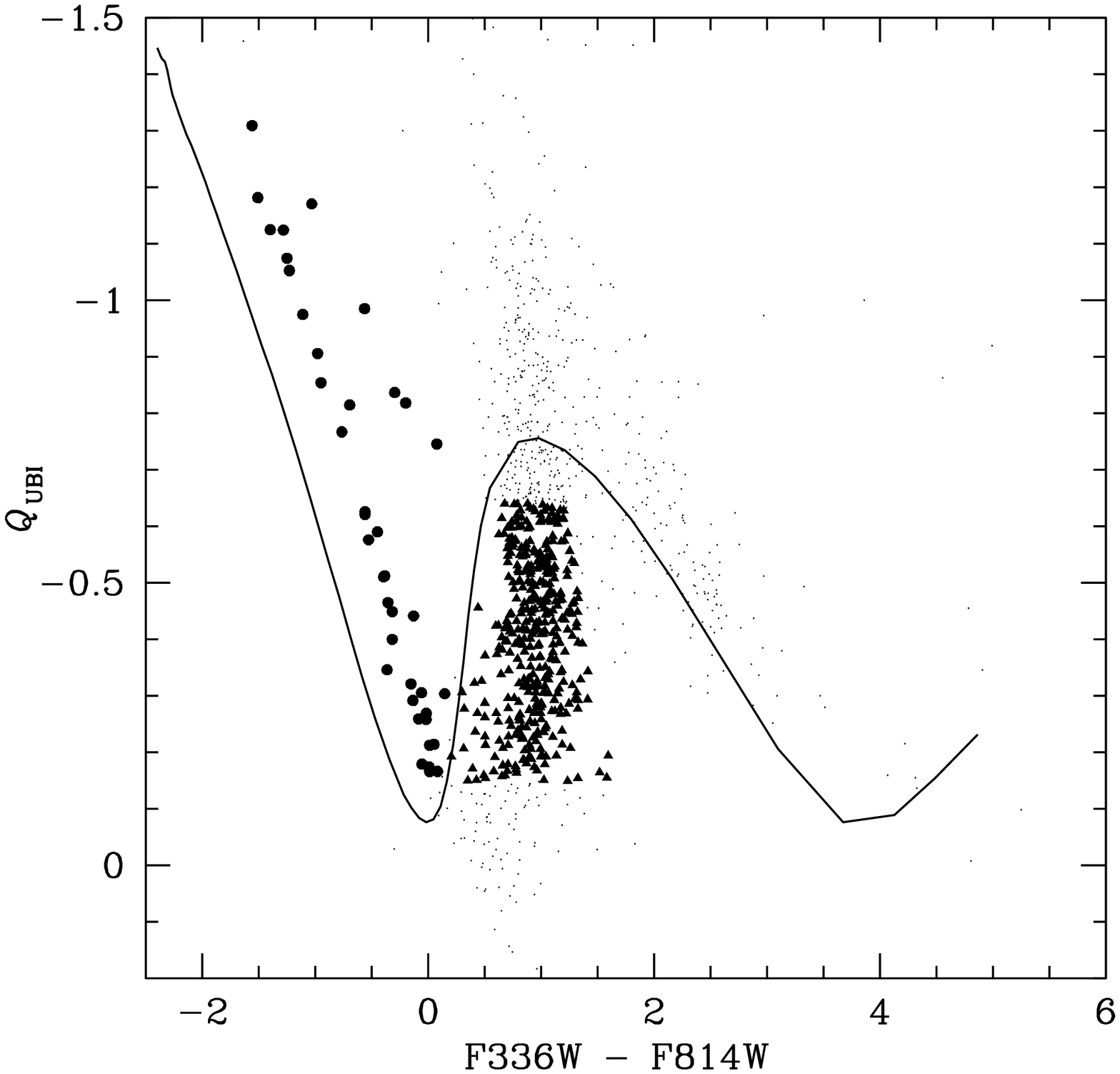}{0.9\linewidth}{0}
\caption{Location of class~I (filled circles), II
  (filled triangles) and III (small dots) stars in the
  \q~vs.~$\bi{F336W}-\bi{F814W}$ plane.\label{fig:qubi_cl}}
\end{figure}

As an example, the result of the division into classes is shown in
Figure~\ref{fig:qubi_cl} for the WF3 chip of the July~1997 observations.
Class~I stars are hotter than about 10,000~K, whereas class~II stars are
between 6,750 and 8,500~K. In  the particular case we are discussing here,
out of a total number of 3,139 stars, 40 belong to class~I (1.3\%), 388 to
class~II (12.4\%) and 615 to class~III (19.6\%). The rest of them
(2096 or 66.8\%)  belong to class IV. 

\placefigure{fig:qubi_cl}

Again, for every star the measured \q\ gives a first order estimate of the
temperature, see Figure~\ref{fig:qubit}, and the excess in \eij{F336W}{F814W}
gives a first order estimate of \ebv\ from equation~(\ref{eq:meth_dui}).
These are, then, used as an input in the second step of the procedure,
the 6-band fit, which we will now describe.

\subsection{Second step: 6-band fit}\label{sec:fit}
In the first step we have used only a subset of the 6 bands available.
Now that the number of possible solutions in the \te-\ebv\ plane
is assessed for every star, we can proceed to fit the entire spectrum
at once with a $\chi^2$ technique using the information gathered in the first
step as a starting point for the fit:

\begin{description}
\item[class I:] The fit is performed in a neighborhood of $\pm0.1$~\mg\
  around the \ebv\ value deducted from the projection in the
  \q~vs.~F336W$-$F814W plane;
\item[class II:] We choose the smallest of the two \ebv\ values from
  the projection described above as the starting point and perform the
  fit in a neighborhood around it, defined as to avoid the other possible
  solutions. This is somewhat arbitrary, but an inspection of
  Figure~\ref{fig:qubi_cl}, for example, does not indicate any compelling
  evidence for a large number of stars with extinction much larger than the
  mean value. Let us stress here that the reddening is still computed
  individually for each star;
\item[class III and IV:] for the various reasons mentioned above, photometry
  is not reliable enough to allow to solve for all of the three unknowns
  simultaneously and the reddening is derived from class~I and II stars.
  We have explored two options: either using the mean value of the 4
  closest class~I and II neighbors or the mean over the entire field.
  In the case of the field of SN1987A the result is the same, as the extinction
  does not show any significant spatial pattern and the local mean is mostly
  equal to the general one. However, when the reddening shows appreciable small
  scale correlations, however, the local mean does give better results, most
  notably a narrower Main Sequence, and the first option should be preferred
  \citep[see, for example, the case of NGC~6822 in][]{bia01}.
\end{description}

The details of the $\chi^2$ fit are described in Appendix~\ref{sec:app_fit}.
While in the selection process we have used constant extinction coefficients
to build \q\ (see equation~(\ref{eq:meth_qubi})), in the 6-band fit they vary
according to the \te\ and \ebv\ values (see Table~\ref{tab:abs} in
Appendix~\ref{sec:app_colors}).

This procedure assures total control over the fitting process. In
fact, even in the case of multiple intersections, it provides a criterion
to choose the most reasonable solution. Once this is done, the 6~band
fit uses  all of the information available and, thus, gives the best
possible answer. Moreover, the whole procedure is very efficient in
terms of computer CPU time usage, since it limits the size of the 
grid in the 6~band fit, which is the most time-consuming part of it
all, only to a small region in the parameter space around the physical
minimum in the \te-\ebv\ plane.

\section{Results of the fit}\label{sec:fresults}
\begin{figure}[!ht]
\efig{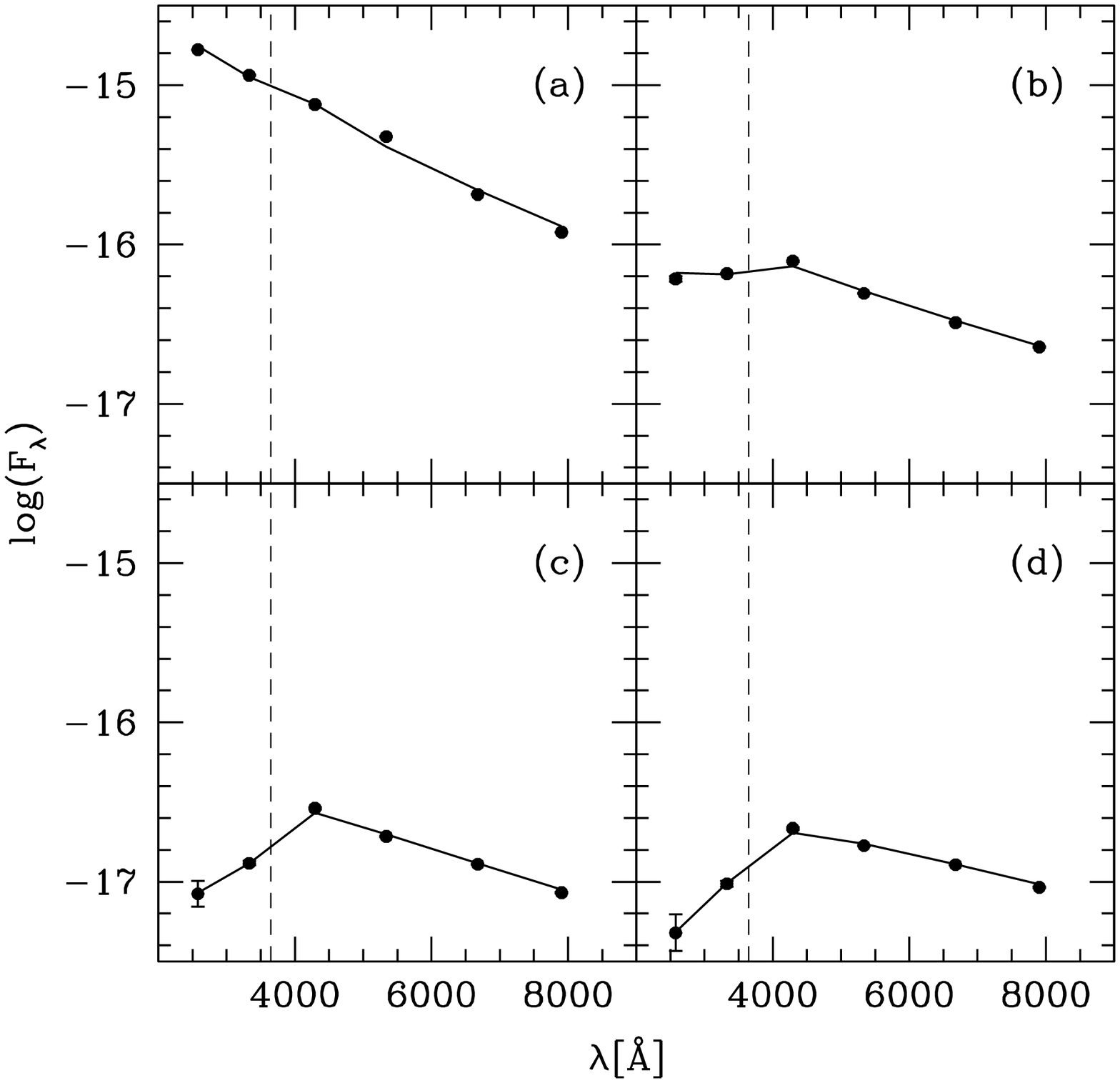}{0.9\linewidth}{0}
\caption{Same as Figure~\protect{\ref{fig:wbsp_nfit}},
  but with the best fitting model shown as a full line. The parameters derived
  from the fit are listed in Table~\protect{\ref{tab:wbsp_fit}}.
  \label{fig:wbsp_fit}}
\end{figure}

As an example of the results of the procedure, in Figure~\ref{fig:wbsp_fit}
we show the spectra of the same four stars of Figure~\ref{fig:wbsp_nfit}
together with the best fitting models. The derived parameters and their
errors are listed in Table~\ref{tab:wbsp_fit}.

\placefigure{fig:wbsp_fit}
\placetable{tab:wbsp_fit}

\begin{figure}[!ht]
\hspace{2cm}\texttt{See 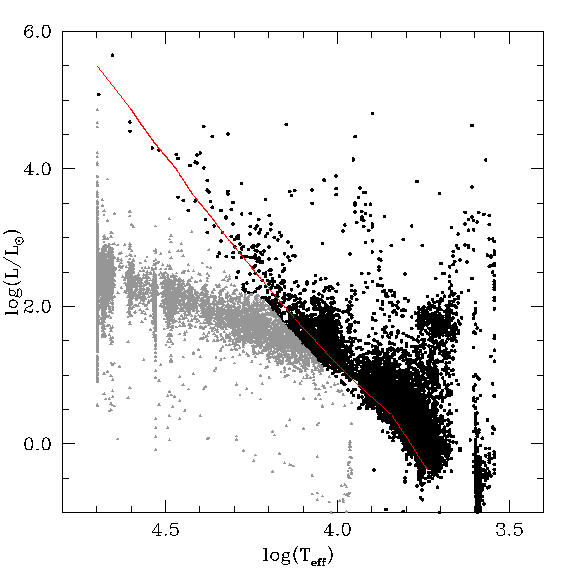}\hfill\efig{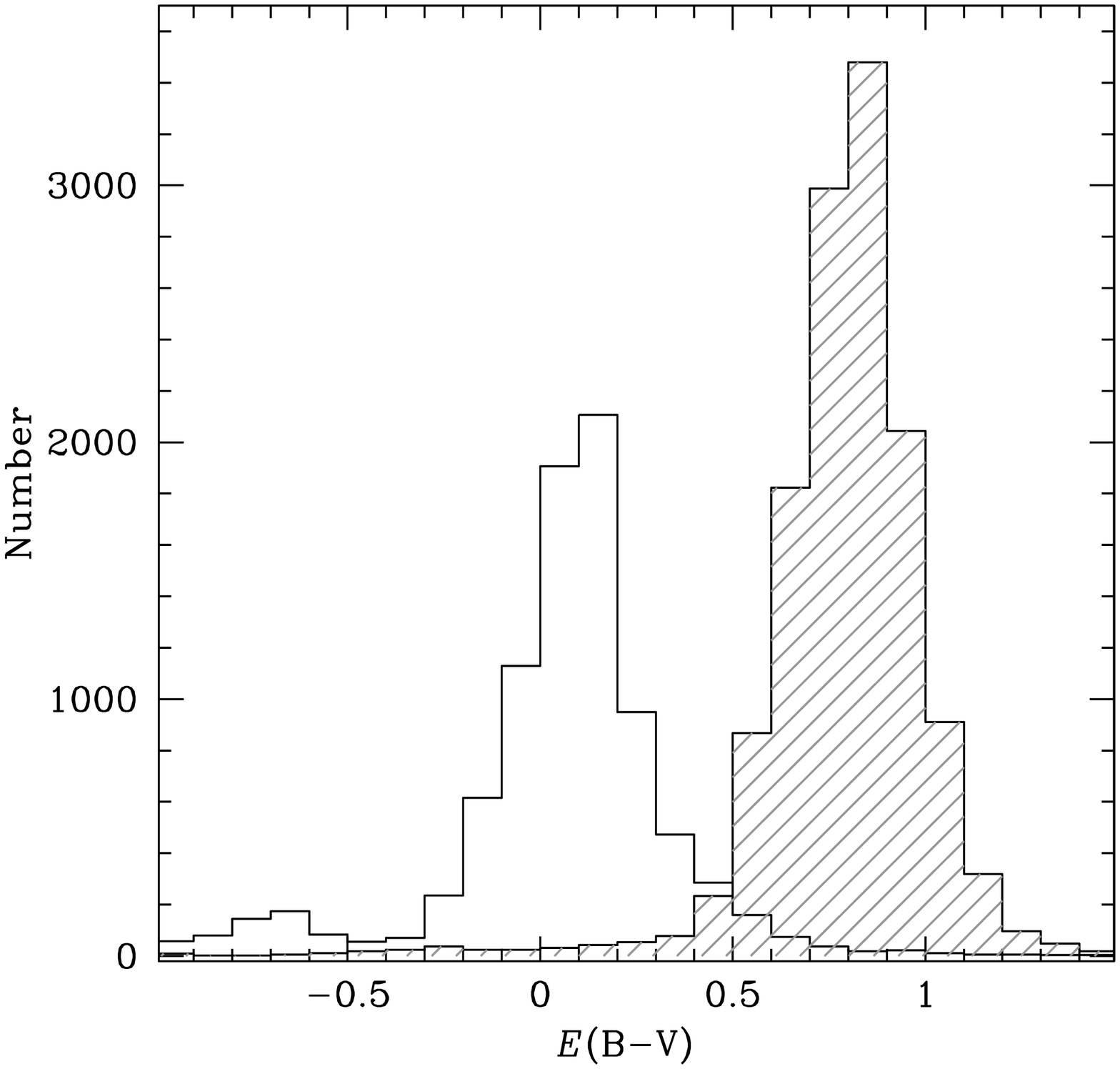}{0.49\linewidth}{0}
\caption{Results of the ``blind'' fit,
  for the stars around SN1987A: the fit described in
  section~\protect{\ref{sec:fit}} is performed directly without the selection
  of section~\protect{\ref{sec:redfree}}. \emph{Left panel:} HR diagram 
  The grey triangles highlight the stars for which the fit returns too high a
  value of \te\ and \ebv. These are stars with multiple solutions for
  which the ``blind'' fit picks the wrong one (see text). The Zero Age Main
  Sequence for Z$=0.3\cdot\mathrm{Z}_\sun$ from the models by
  \protect{\citet{bc93}} and \protect{\citet{ccs94}} for masses below
  25~M$_\sun$ and by \citet{sch93} above 25~M$_\sun$ is shown as a full line.
  \emph{Right panel:} \ebv\ distribution resulting from the ``blind'' fit.
  The high reddening peak (hatched histogram) is entirely made of stars
  plotted in grey in the left panel, thus confirming that their location
  in the HR diagram is due to an erroneous reddening determination.
  \label{fig:blind}}
\end{figure}

The HR diagram for the 21,955 stars in the field of SN1987A is displayed in
Figure~\ref{fig:sn87a_hrms}. It is interesting to compare it to what one
would have obtained from a ``blind'' fit of the spectrum, \ie by skipping the
selection described in section~\ref{sec:redfree} and performing directly the
6-band fit of section~\ref{sec:fit}. This is shown in the left panel of
Figure~\ref{fig:blind}.

\placefigure{fig:sn87a_hrms}
\placefigure{fig:blind}

The fit for the stars marked in grey in the left panel of
Figure~\ref{fig:blind} is obviously wrong as they occupy an impossible
region of the HR diagram. An inspection of the right panel of the same
Figure reveals that all of these stars, and only them, were fitted with
too high a value of \ebv\ and, consequently, of \te. These are all
stars that fall in regions $C$ or $D$ in Figure~\ref{fig:qubit} for
which the broad band spectrum can be interpreted either as a heavily
reddened hot star or as a cold one with little dust in front of it. The
fit just selects the solution with the lowest $\chi^2$, even if it is
marginally lower than other possible minima. Which one of the
possible solution  has the lowest $\chi^2$ depends on such things as
the \ebv\ and \te\ grids used in the fit and, in the case shown in
Figure~\ref{fig:blind}, the wrong one is chosen for two stars out of
three. For many of these stars the available photometry is limited to F439W
and redder filters. As we will see in section~\ref{sec:need}, then, the data
are very little sensitive to temperature and these cold stars can be mistaken
for objects as hot as 50,000~K.

On the other hand, selecting the right solution \emph{before} performing the
fit eliminates the problem almost completely. There still are roughly 600
stars (less than 3\% of the total) under the Zero Age Main Sequence locus.
However, they are mostly class IV objects, \ie stars with very poor photometry,
or stars with peculiar colors, \ie the T~Tauri stars described in detail
by \citet{prsk00} or unresolved binaries.

The stars with multiple solutions are the late-type ones. Being
able to measure \emph{directly} the reddening for them is important
for two reasons.
Firstly, as low mass stars are much more numerous than higher mass
ones, the number of reddening determinations increases dramatically. In
fact the late type stars for which  eventually  we are able to
measure the reddening are typically a factor of 10 more numerous than
those earlier that A0. Secondly, stars of different masses have
different lifetimes and there is no a priori reason why different
generations of stars should be affected by the same amount of
extinction \citep[see][for an example in the LMC itself]{zar99}. By
extending our analysis to a wide range of masses we can check for
population-dependent effects and deredden the different generations of
stars in the most appropriate way.

\begin{figure}[!hb]
\dfig{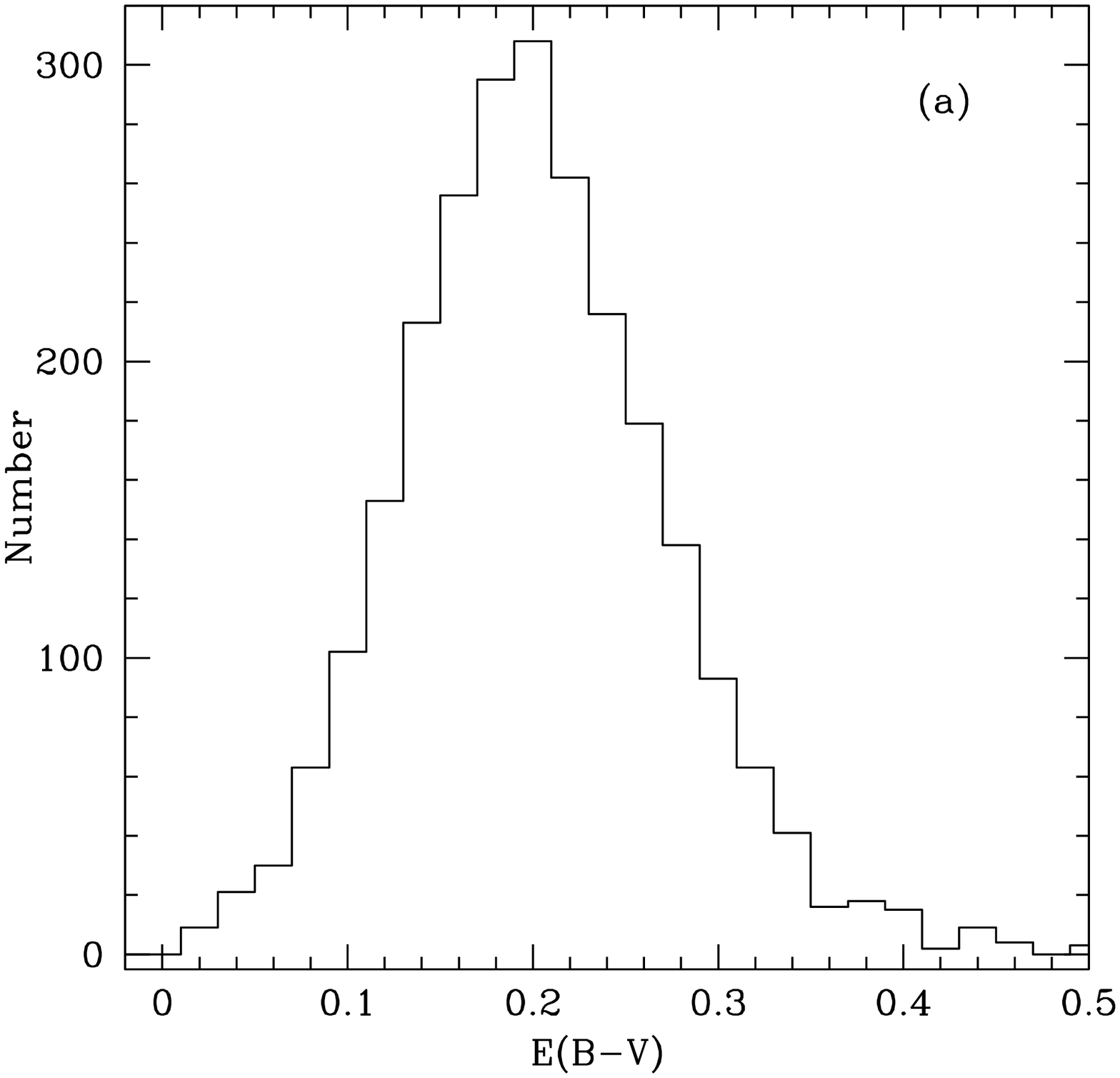}{0.49}{0}{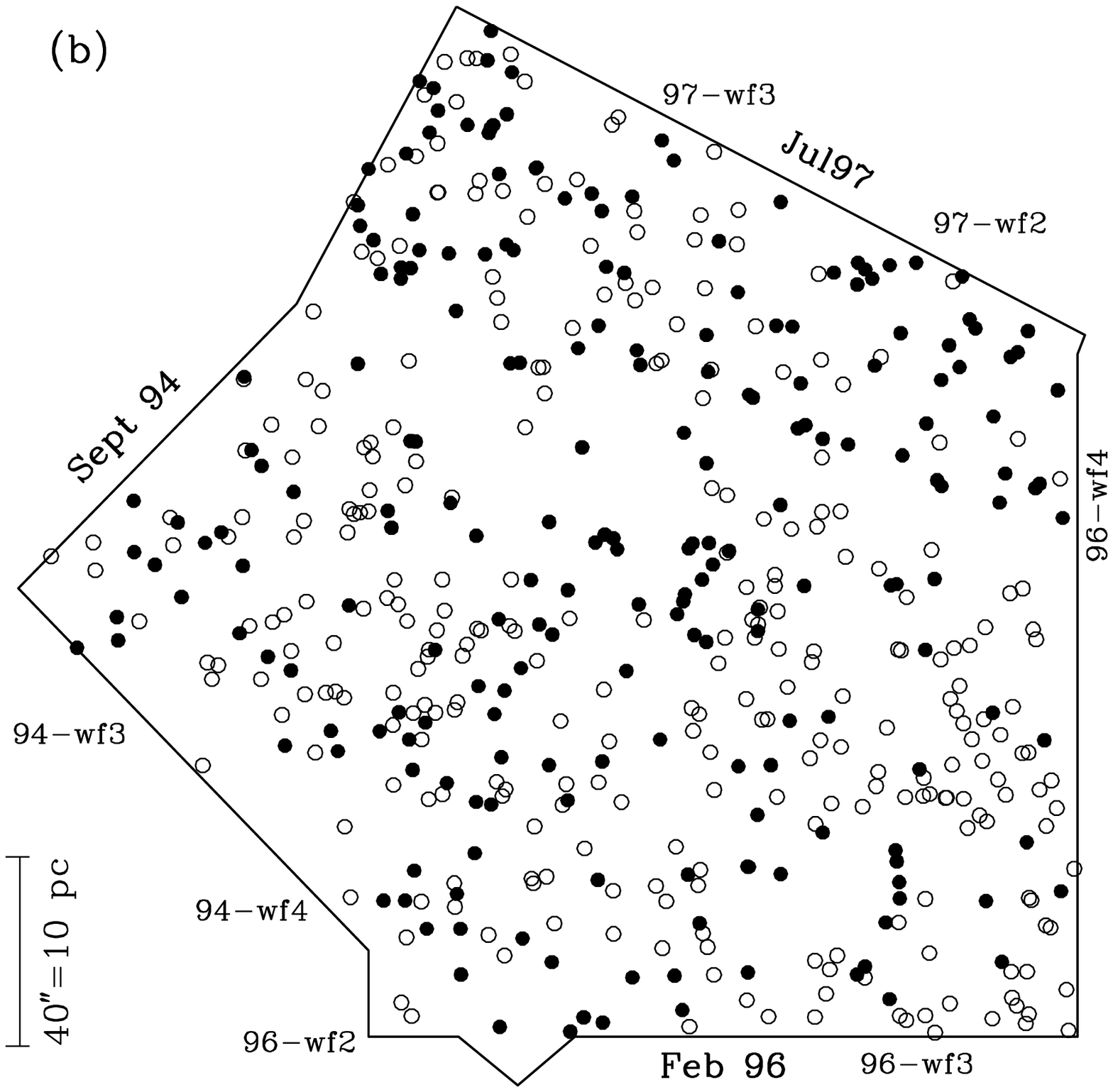}{0.49}{0}
\caption{Panel~(a): reddening
  histogram from the stars in the field of Supernova~1987A. Only stars for
  which \ebv\ was measured individually (the 2,510 class~I and~II stars, see
  the text) are used here. Panel~(b): spatial distribution of stars with high
  ($\ebv>0.3$, filled circles) and low ($\ebv<0.1$, open circles) reddening.
  \label{fig:sn87a_ebv}}
\end{figure}

In addition to measuring the intrinsic properties of the stars shown in
Figure~\ref{fig:sn87a_hrms}, our procedure also provides individual \ebv\
measurements for more than 2,500 stars, on average one every 13 square
arcseconds. The resulting histogram and spatial distribution of interstellar
reddening are shown in Figure~\ref{fig:sn87a_ebv} \citep[see also][]{prsk00}.
As it can be seen, the initial suspicion is confirmed and the reddening is
indeed very patchy and varies on scales of a few arcseconds.

\placefigure{fig:sn87a_ebv}

Besides providing a very accurate reddening map, measuring individual
values of \ebv\ also leads to a narrowing of the various features in
the Color-Magnitude diagram, in particular the Main Sequence. This is
of great help when interpreting the data, because it allows one to
disentangle the broadening due to differential extinction from the one
caused, for example, by the overlap of several generations of stars or
by the presence of binary stars.

\begin{figure}[!hb]
\efig{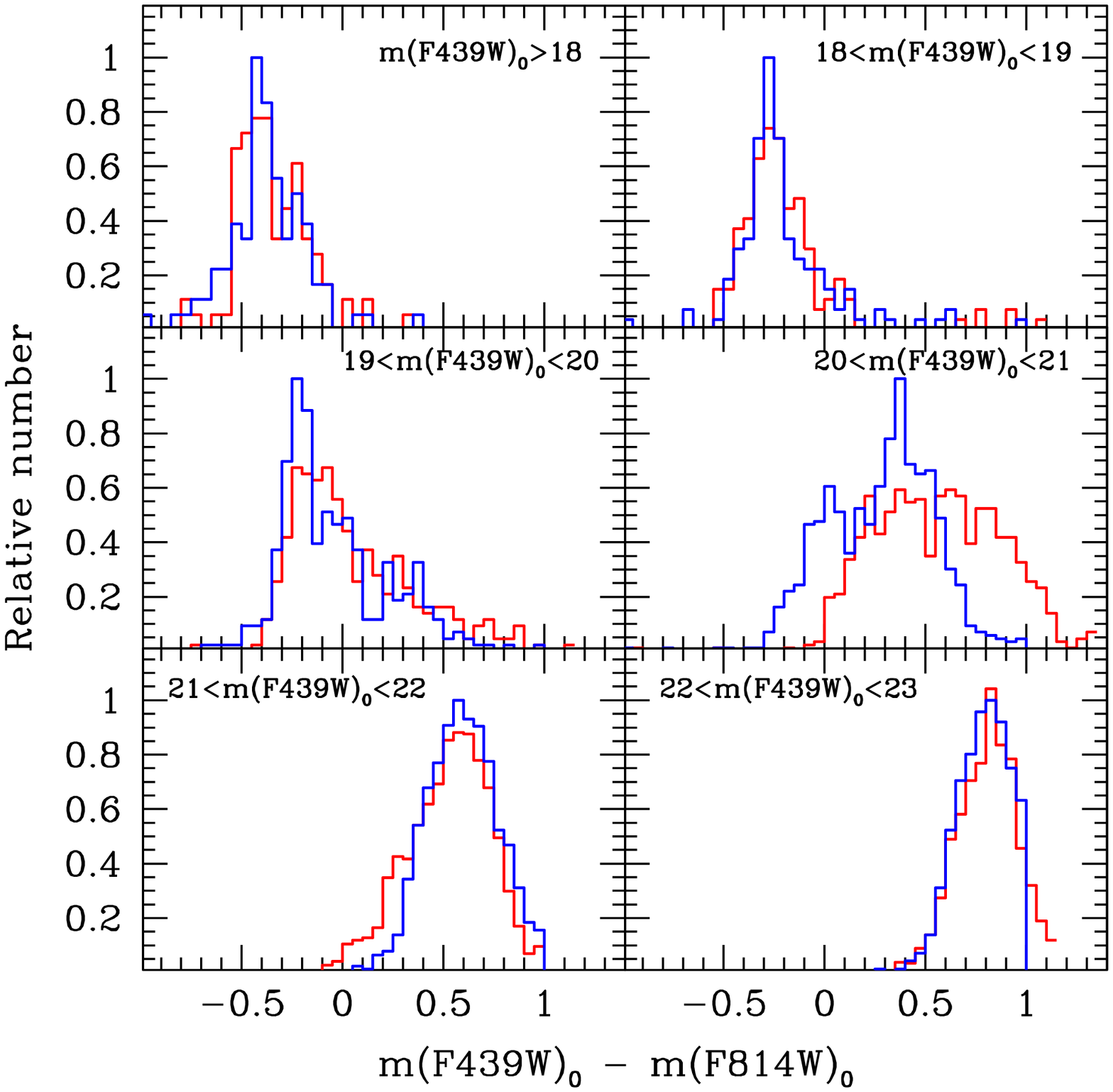}{0.9\linewidth}{0}
\caption{Width of the Main Sequence as a function of
  magnitude before (red line) and after (blue line) reddening correction.
  The magnitude interval along the Main Sequence is indicated in each panel.
  The faintest magnitude bin heavily suffers from incompleteness. Also, the
  raw histograms were shifted in color in order to have the same mode as the
  corrected ones. As it can be seen, the corrected histograms are
  systematically narrower than the uncorrected ones.\label{fig:beaft_msh}}
\end{figure}

In Figure~\ref{fig:beaft_bihall} we compare the stars in the
F439W~vs~(F439W$-$F814) diagram before and after reddening correction. The
density contours in panels~(a) and~(b) are spaced by factors of 2. Apart from
the obvious fact that the corrected stars are bluer and more luminous, it is
apparent that the features in the reddening corrected diagram are narrower
than the corresponding ones in the observed CMD. In order to quantify this
effect let us consider a color cut through the Main Sequence, like the one
displayed in panel~(c) of Figure~\ref{fig:beaft_bihall} (see also
Figure~\ref{fig:beaft_msh}). Here, the blue line histogram represents the
color distribution of the dereddened Main Sequence stars with $\dfive<0.1$ in
the $21>$m(F439W)$_0>22$ magnitude interval. The red line histogram gives the
distribution of the same stars before reddening correction. To facilitate the
comparison, this latter is shifted in color so as to have the same mode as the
dereddened one. As apparent, the original distribution is definitely more
spread out in color: its rms is 24\% larger than the corrected one (0.21~\mg\
versus 0.17). This is the case all along the Main Sequence, as shown in
Figure~\ref{fig:beaft_msh}, where we plot the color distribution for
all the stars with $\dfive<0.1$ in different magnitude bins. The
dereddened histograms (blue line) always display a smaller scatter than
the observed ones (red line), confirming the effectiveness of the
correction procedure.

\placefigure{fig:beaft_bihall}
\placefigure{fig:beaft_msh}

As it can be expected, the color distribution is even narrower when
only the stars with individual reddening determinations, \ie those we
have classified as class~I and II, are considered. This is shown in
Figure~\ref{fig:beaft_bihind}. In this case the rms of the
reddening-corrected Main Sequence histogram, again in the
$21>$m(F439W)$_0>22$ magnitude interval, is  0.14~\mg, while the one of
the same stars before correction is 0.18~\mg, \ie 29\% larger. A simple
Kolmogorov-Smirnov test shows that the widths of the two distributions
are different at the 99.9995\% level.

\placefigure{fig:beaft_bihind}

To conclude, let us remark again that Class~III and IV stars were not
corrected with and individually determined value of \ebv, but, rather,
with the mean value from 2,510 Class~I and II neighbors. It is important to
realize that, given the large number of these stars, the mean reddening
used to correct  Class~III and IV stars is on average very accurate. Of
course, it is still possible that a few class III and IV stars have, in
reality, extinction values significantly different from their local mean,
but the global effect is negligible.

The full discussion of the stellar content of the region around SN1987A
can be found in \citet{rom98} and \citet{prsk00}.

\section{The role of gravity and metallicity}\label{sec:meth_gte}
So far we have used model atmospheres  only with one  surface gravity
($\log(g)=4.5$, appropriate for Main Sequence stars) and metallicity
(Z$=0.3\cdot\mathrm{Z}_\sun$, the expected mean value for the LMC).
Even though these are sensible assumptions for the environment under study,
we have to verify their influence on the results we derive.

\begin{figure}[!hb]
\efig{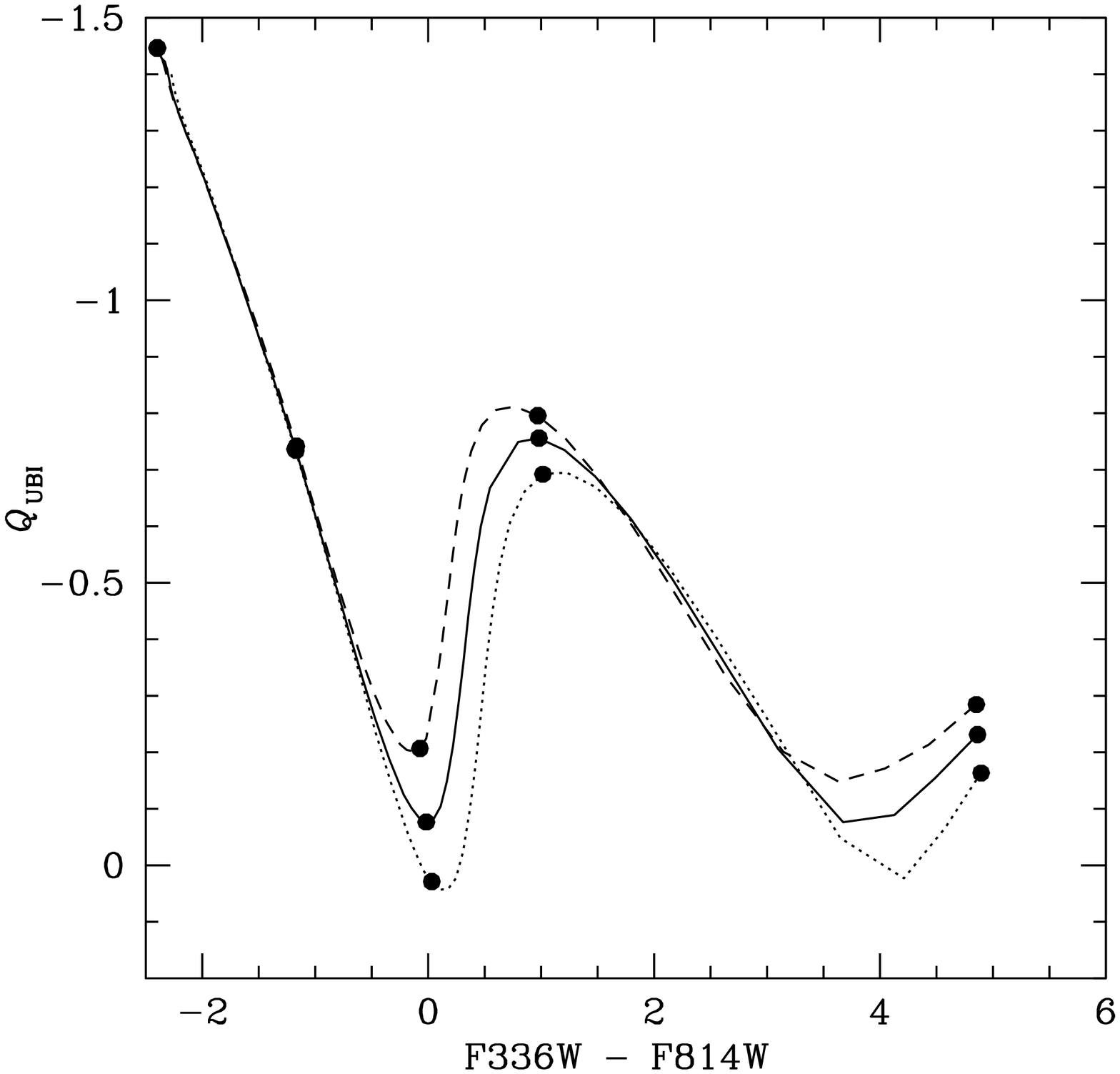}{0.9\linewidth}{0}
\caption{Zero-reddening \q~vs.~$\bi{F336W}-\bi{F814W}$
  relation for three values of surface gravity: $\log(g)=5$ (dashed line),
  $\log(g)=4.5$ (solid line) and $\log(g)=4$ (dotted line). The model
  atmospheres are those of \protect{\citet{bes98}} for
  Z$=0.3\cdot\mathrm{Z}_\sun$. The points for \te=50,000, 15,000, 9,250, 6,000
  and 3,500~K are marked on each curve.\label{fig:meth_qubig}}
\end{figure}

Let us start by considering the surface gravity. The theoretical zero-reddening
locus in the \q~vs.~F336W$-$F814W plane is shown in
Figure~\ref{fig:meth_qubig} for three values of surface gravity, namely
$\log(g)=4$ (dotted line), 4.5 (solid line) and 5 (dashed line). As it can be
seen, the three curves
overlap perfectly for F336W$-$F814W$\lesssim -0.5$, \ie $\te\gtrsim
11,000$~K and F336W$-$F814W$\gtrsim 1.5$, \ie $\te\lesssim 5,500$~K.
Elsewhere, however, different surface gravities lead to different
expected zero-reddening loci in the \q~vs.~F336W$-$F814W diagram. In
particular, this influences the division in classes that is the
foundation of the method described above.

\placefigure{fig:meth_qubig}

However, as it can be seen, for example, in
Figure~\ref{fig:beaft_bihind}, only stars on or near the Main Sequence
are used to determine the reddening. Adopting the stellar evolutionary
models by \citet{bc93} and \citet{ccs94}, we see that for
Z$=0.3\cdot\mathrm{Z}_\sun$ the gravity is nearly constant along the
Main Sequence and that its value is $\log(g)=4.5$, the value we have
used in the fit. This is a sound assumption over a large range of
metallicities because extensive model calculations show that surface
gravity variations as a function of chemical composition are quite small.
For example, at $(\bi{F336W}-\bi{F814W})_0=0$ the surface gravity ranges from
$\log(g)\simeq 4.4$ for a solar metallicity star to $\log(g)\simeq 4.6$ for
Z=Z$_\sun/20$. Considering that for every metallicity the gravity is
almost constant along the Main Sequence one can safely adopt the value
of $\log(g)=4.5$, independent of chemical composition.

Once a star evolves off the Main Sequence its surface gravity changes
appreciably.
For example, $\log(g)\simeq 3$ is a typical value for the stars in the
Red Giant clump, \ie the stars at $\log(\te)\simeq 3.7$ and
$\log(\mathrm{L}/\mathrm{L}_\sun)\simeq 1.75$ in
Figure~\ref{fig:sn87a_hrms}. Even though the \q~vs.~F336W$-$F814W
relation is, indeed, sensitive to gravity, the result of our global fit
is not. This is because, as we have seen, we do not fit simultaneously
temperature and reddening for stars off the Main Sequence, but rather
we assign to them the mean \ebv\ value of their class I and II
neighbors, which are on the Main Sequence and, hence, have
$\log(g)\simeq 4.5$. For stars that are off the Main Sequence, we just
perform the fit to solve for the radius and effective temperature. Once
the reddening is correctly determined, the result of the multi-band fit
is no longer sensitive to the surface gravity adopted in computing the
theoretical colors. We have checked this by dividing the stars into
classes using the \q~vs.~$\bi{F336W}-\bi{F814W}$ relation for $\log(g)=4.5$,
and then fitting the stars with colors computed for various values of
the gravity. The maximum variations in temperature for the stars on the
Red Giant branch are of 3\%, \ie much smaller than the error deriving
from the fit itself, and with no systematic trend with gravity. Given
this result, we have decided to use $\log(g)=4.5$ all throughout our
analysis.

\begin{figure}
\efig{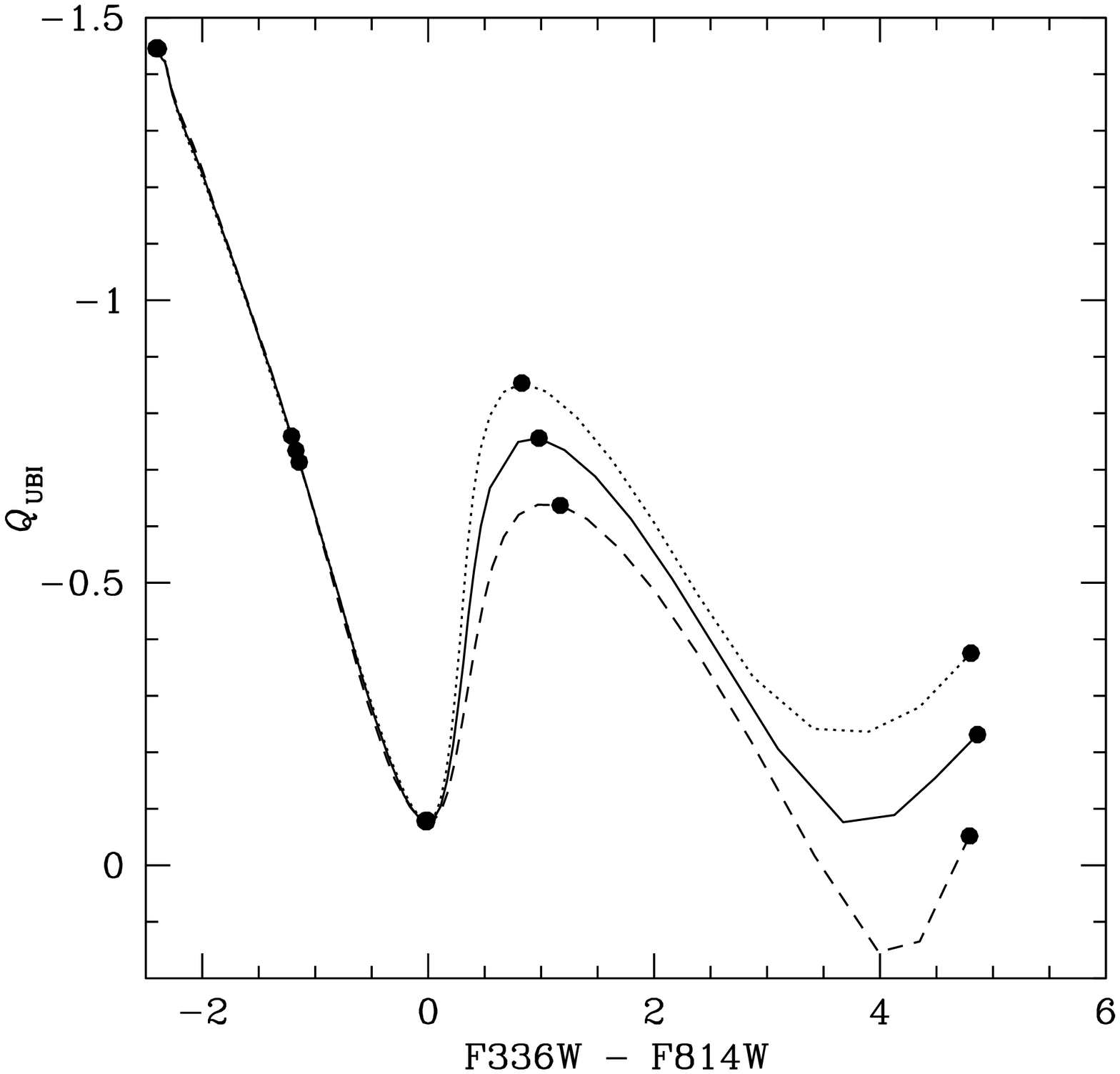}{0.9\linewidth}{0}
\caption{Zero-reddening \q~vs.~$\bi{F336W}-\bi{F814W}$
  relation for three values of metallicity: Z=Z$_\sun$ (dashed line),
  Z$=0.3\cdot\mathrm{Z}_\sun$ (solid line) and Z=Z$_\sun/10$ (dotted line). The
  model atmospheres are those of \protect{\citet{bes98}} for $\log(g)=4.5$.
  The points for \te=50,000, 15,000, 9,250, 6,000 and 3,500~K are marked on
  each curve.\label{fig:meth_qubiz}}
\end{figure}

In addition to influencing the stellar structure, the chemical
composition also plays a role in determining the star's spectrum,
through the atmospheric opacity. As a consequence, the colors also
depend on this parameter. The \q~vs.~$\bi{F336W}-\bi{F814W}$ relation for
different values of Z is shown in Figure~\ref{fig:meth_qubiz}. Again,
as it can be seen, the variations are negligible for the stars for
which  the reddening is an individual fit parameter, \ie Class I and
II stars. 

\placefigure{fig:meth_qubiz}

Also in this case, then, once the subdivision into classes is made and the 
local mean reddening for Class III and IV stars is determined from the
individual values of Class I and II neighbors, the result of the multi-band
fit is insensitive to the details of the stellar atmospheres. The comparison
of the fits performed with colors computed for different metallicities shows
that the derived parameters do not depend appreciably on metallicity.

We conclude that in applying our fitting procedure, as described
in section~\ref{sec:proc}, one can safely use  colors computed with
models by \citet{bes98} for Z$=0.3\cdot\mathrm{Z}_\sun$ and
$\log(g)=4.5$ to derive accurate stellar parameters and 
interstellar absorptions for all stars.

\section{The bands needed}\label{sec:need}
For every star there are three unknown quantities: the angular radius,
the effective temperature and the reddening. Therefore, for each
star one  needs  at least data in three bands to make the problem
mathematically well defined. For good results the measurements in those
three bands must also be rather accurate, thus requiring that the bands
be selected so as to fall around the wavelength of the maximum of the
spectral distribution of any given star, and have a suitable wavelength
spacing to assure a sufficient baseline for the spectral shape
analysis. In practice this is achieved with bands separated from each
other by about 30-40\% of their central wavelengths.   Moreover, when
dealing with a {\it stellar  population} one wants to study stars
within a wide range of temperatures, and, therefore, more than three
bands are absolutely needed. In this section we will investigate how many
and which bands are needed to constrain the fit.

\begin{figure}[!ht]
\dfig{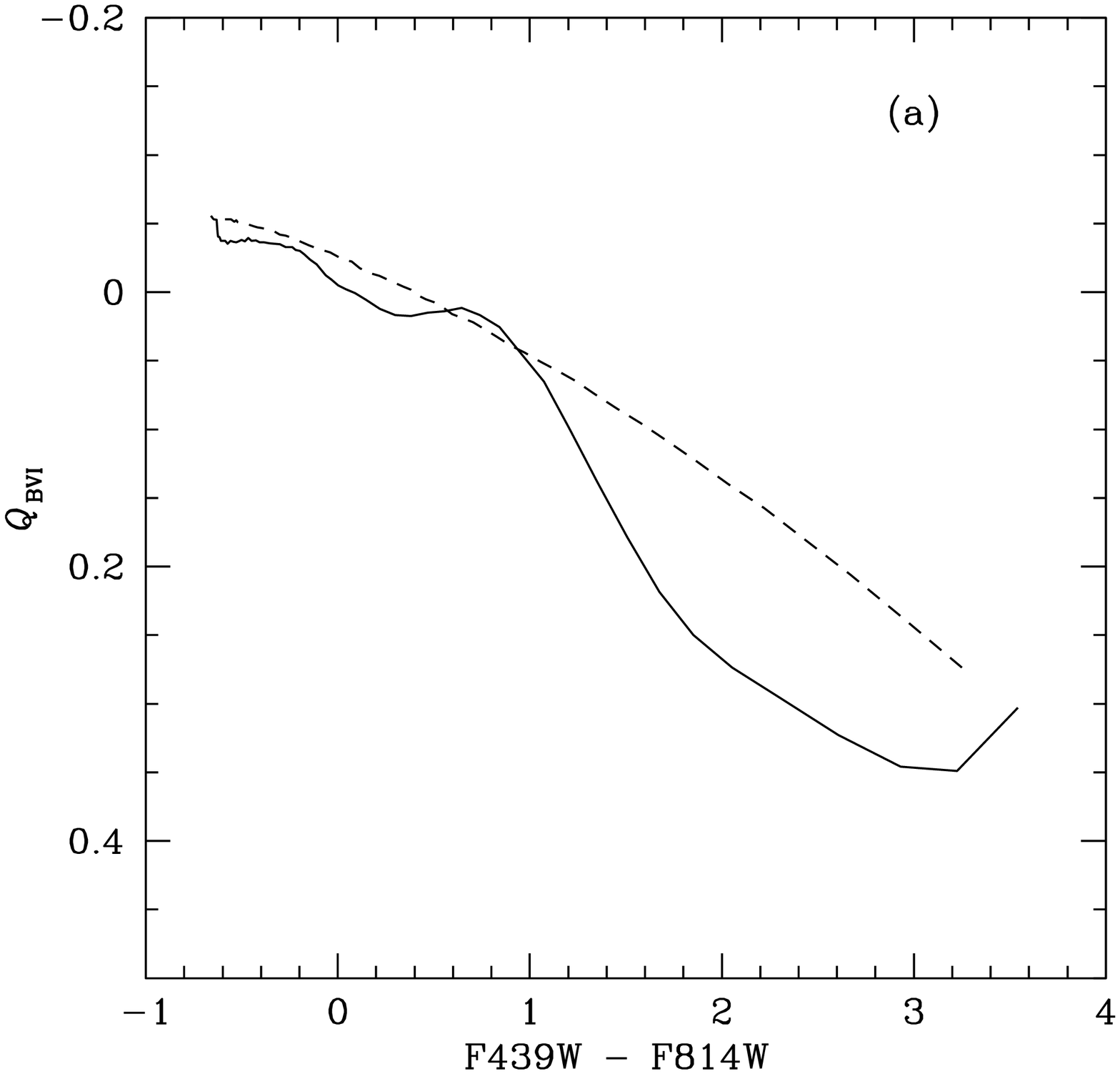}{0.49}{0}{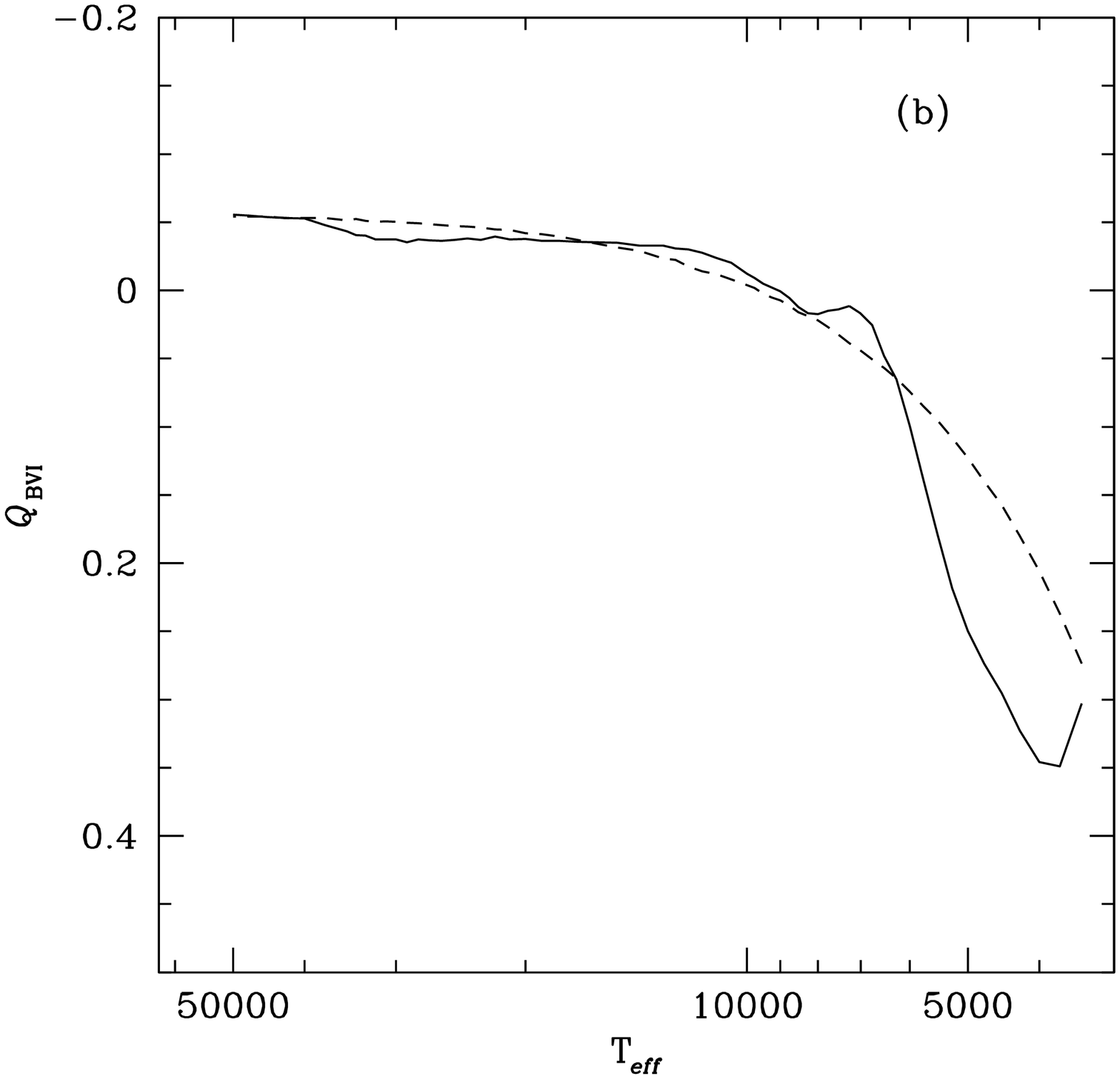}{0.49}{0}
\caption{The $Q_\mathrm{BVI}$ reddening-free
  color as a function of $\bi{F439W}-\bi{F814W}$ (panel~(a)) and \te\
  (panel~(b)) for the \protect{\citet{bes98}} model atmospheres for
  Z$=0.3\cdot$Z$_\sun$ and $log(g)=4.5$. The case of black body spectra is
  shown in both panels as a dashed line.\label{fig:qbvi_bb}}
\end{figure}

Let us start by noticing that a U-like band is mandatory to recover both \te\ 
and \ebv. To illustrate this point let us define the reddening-free color
$Q_{\mathrm{BVI}}$ combination of B, V and I. By analogy with
equation~(\ref{eq:meth_qubi}):

\begin{equation}
Q_\mathrm{BVI}\equiv\left(\bi{B}-\bi{V}\right)-\frac{E(\bi{B}-\bi{V})}
{E(\bi{V}-\bi{I})}\cdot \left(\bi{V}-\bi{I}\right)
\label{eq:meth_qbvi}
\end{equation}
$Q_\mathrm{BVI}$ is shown in Figure~\ref{fig:qbvi_bb} as a function of
$\bi{F439W}-\bi{F814W}$ in panel~(a) and \te\ in panel~(b).

\placefigure{fig:qbvi_bb}

Since the the B, V and I filters do not include the Balmer jump, the relation
between $Q_{\mathrm{BVI}}$ and B$-I$ is monotonic\footnote{The hook at very
low temperatures is due to the red leak in the \wfpc\ F439W filter.},
thus solving the problem of ambiguities in the solution. However, one
immediately notices that the slope of the curve is much shallower than
in the \q\ case. In fact, the total excursion of $Q_\mathrm{BVI}$
between 3,500 and 50,000~K is only 0.4~\mg\ and it is almost totally
insensitive to temperatures higher than 6,000~K, \ie
$\bi{F439W}-\bi{F814W}\lesssim~1$.

\begin{figure}
\hspace{-0.25cm}\epsfig{file=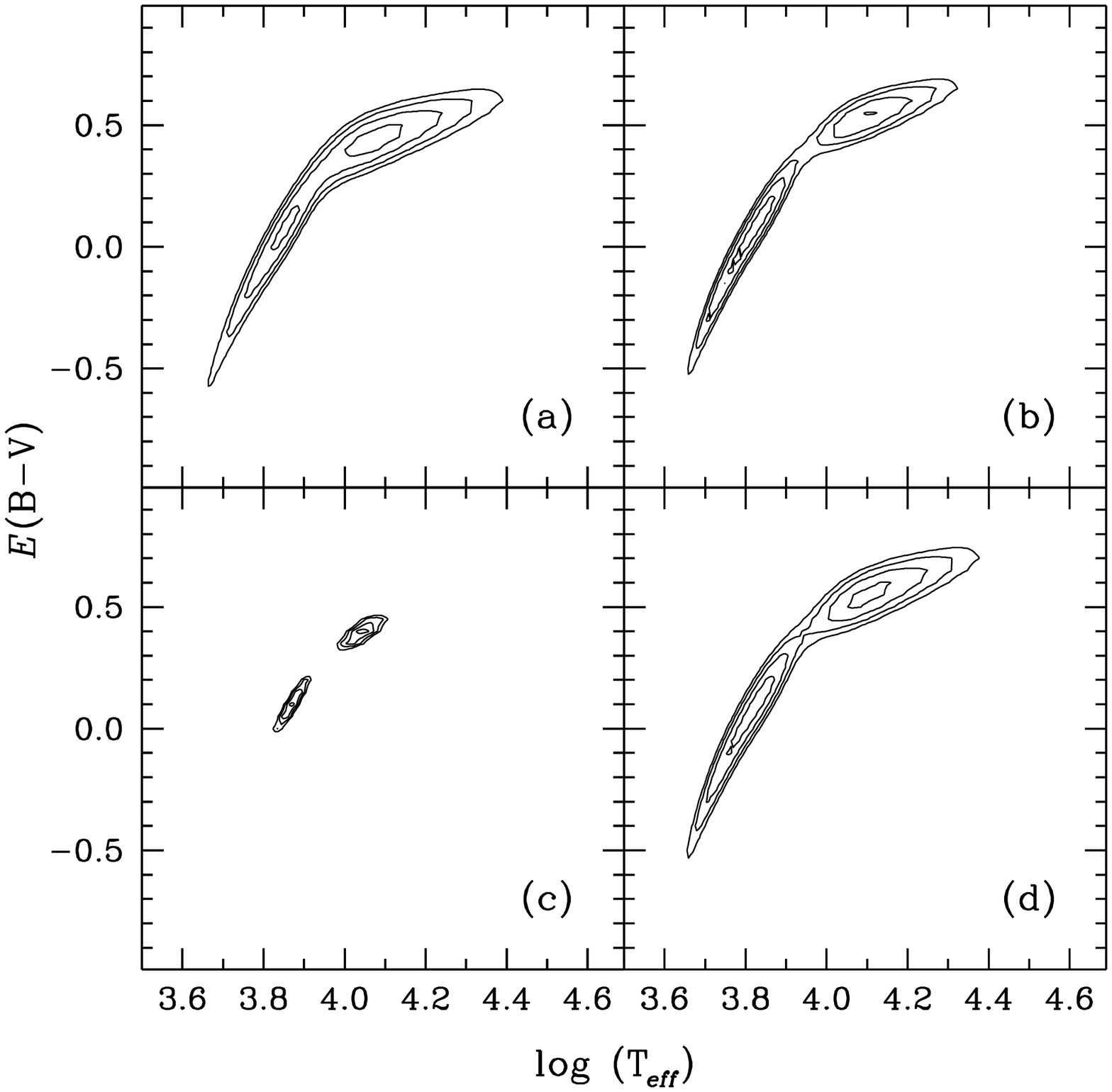, height=0.49\linewidth}\hfill\texttt{See 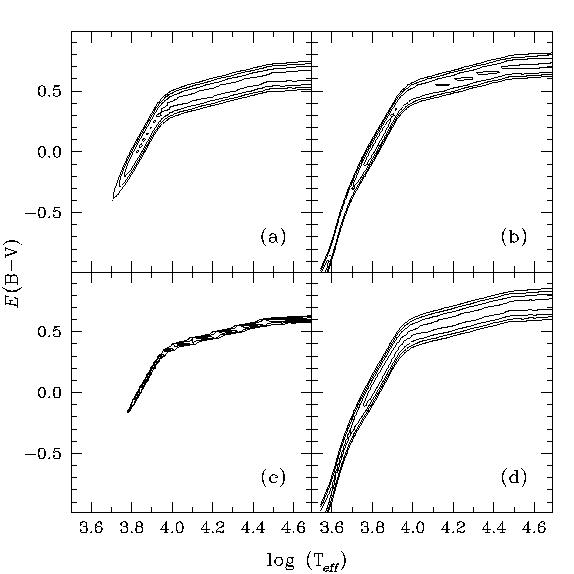}\hspace{3cm}
\caption{Same as
  Figure~\protect{\ref{fig:mchi_I}}, but for stars with multiple solutions.
  \label{fig:mchi_II}}
\end{figure}

The combined effect of the weak dependence of $Q_\mathrm{BVI}$ on temperature
and of photometric errors is to make every combination \ebv-\te\ along the
reddening vector essentially equivalent. This is shown in
Figure~\ref{fig:mchi_I} and Figure~\ref{fig:mchi_II} where $\chi^2$ maps
of the fit using only four bands (F439W, F555W, F675W, and F814W, left panel)
are compared to those using all six of them (right panel). As it can be seen,
when only 4 bands are used the $\chi^2$ has no distinct minima, and, hence,
it is not possible to determine both the reddening and the temperature neither
for hot stars for which the solution is unambiguous (Figure~\ref{fig:mchi_I})
nor for the cold ones with multiple minima (Figure~\ref{fig:mchi_II}).

\placefigure{fig:mchi_I}
\placefigure{fig:mchi_II}

\begin{figure}
\efig{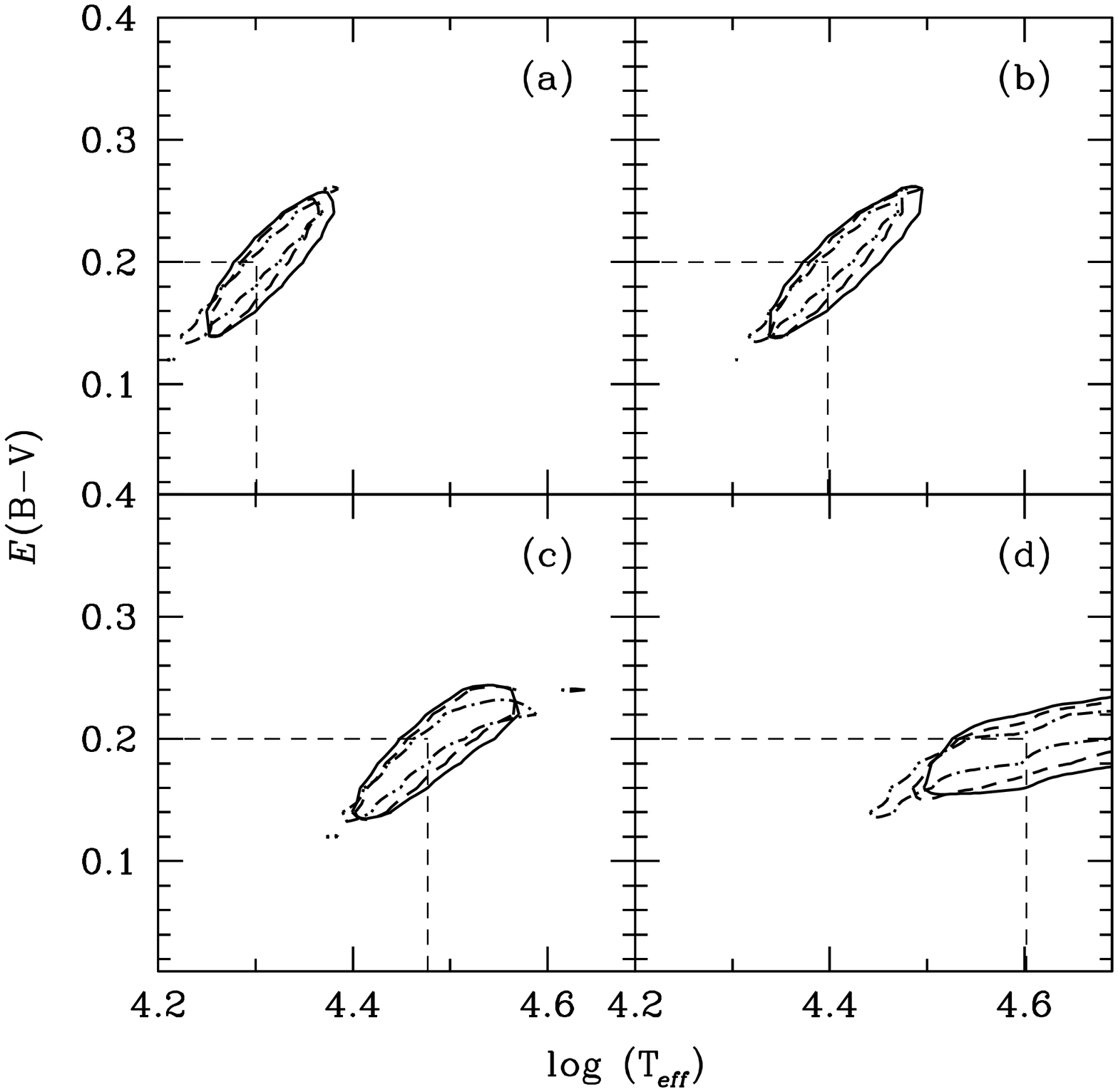}{0.9\linewidth}{0}
\caption{$\chi^2$ contour map for artificial stars with
  $\ebv=0.2$ and $\te=20,000~\mathrm{K~(panel~(a))}$, 25,000~K~(b),
  30,000~K~(c) and 40,000~K~(d). The contours correspond to the $1\sigma$
  confidence level for a 5-band fit. In all cases the same 4 optical filters
  are used (F439W (B), F555W (V), F675W (R) and F814W (I)) in combination
  with different UV passbands: F170W (dot-dashed line), F255W (dashed line) or
  F336W (full line). The photometric error is assumed to be the same in all 3
  UV filters. The thin straight lines mark the input parameters of the model
  stars.\label{fig:fit-uv}}
\end{figure}

Also, let us notice here that, while observations at wavelengths
shorter than the Balmer jump are mandatory to recover both \te\ and
\ebv, it is preferable not to go too far in the ultraviolet, unless
one can afford to make observations in more than {\it one} UV band.
Qualitatively, the reason for this is that shorter wavelengths are
more affected by interstellar reddening and small differences in
\ebv\ along different lines of sight may result in large differences 
in the observed flux, thus making the ultraviolet flux of a moderately
reddened hot star almost indistinguishable from the one of an
unreddened, colder one. Quantitatively, the situation is illustrated in
Figure~\ref{fig:fit-uv}, where we plot the $1\sigma$ contour levels for
artificial stars of different effective temperature as observed through
4 optical filters (F439W, F555W, F675W and F814W) and one UV one: F170W
(dot-dashed line), F255W (dashed line) or F336W (full line). The contours,
computed assuming the same photometric error in all 3 UV filters, overlap
almost perfectly. This means that one does not gain in accuracy by observing
at shorter wavelengths, even for hot stars. Since the sensitivity of
normal CCDs and the emission of cold stars both drop dramatically the
further one moves to the ultraviolet, a filter in the region of the U
band is the ideal choice for the filter bluer of the Balmer jump.

\placefigure{fig:fit-uv}

As we have pointed out at the beginning of this section, there are 3
unknowns in the fit so that 6 bands are actually redundant and 4 could
be enough to cover the wide range of temperatures spanned by stars
in regions with mixed stellar populations. In fact, the method works 
rather well  also with only 4 bands, provided that they encompass the
Balmer jump. This is shown in Figure~\ref{fig:sn87a_parf_hr}, where the
HR diagram for the control field located $8^\prime$ south-east of
SN1987A, derived following our procedure, is shown. In this case, as
reported in Table~\ref{tab:sn87a_parf_log}, the four available
photometric bands were F300W, F450W, F675W and F814W. The star symbol
shows the location of the brightest star in the \wfpc\ field of view.
It is so heavily saturated in the  \wfpc\ chip that its image was
spilling over several neighboring stars thus making impossible to make
direct measurements, and, therefore, its photometry is taken from
\citet{fitz88}. 

\placefigure{fig:sn87a_parf_hr}

As it can be seen, the observed Main Sequence, although still in very
good agreement with the theoretical one, is less sharply defined than
the one in the field of SN1987A (see Figure~\ref{fig:sn87a_hrms}) and  
more stars fall under the theoretical ZAMS. 
Using a reduced set of filters has affected the result in two ways.
Firstly, there is only one  band   blueward of the Balmer jump (F300W) 
instead of two (F255W and F336W): the fit is less sensitive to
temperature for hot stars and, hence, for them, the results are less
accurate. Secondly, since the observations were taken as ``parallels"
to FOS primaries, we did not have enough time to make observations both
in the F555W, or a similar band like F606W, and in the F675W filter.
We  opted for this latter one to identify stars with \ha\ emission more
accurately. As a result, though, the photometric points are not spaced
in wavelength quite evenly. As a consequence, the fit for the cooler
stars, the ones that are not well exposed in F300W and F450W bands, has
to rely on a short wavelength baseline and, thus, is less precise than
it would have been if observations at roughly 5000~\AA\ had been
available. This also results in a larger fraction of stars below the Main
Sequence (roughly 600 out of a total of nearly 13,100, \ie 4.6\%) when
compared to the main field (roughly 600 out of almost 22,000, \ie less than
3\%). \emph{In general, however, the  resulting stellar parameters are still
of very good quality}, with more than 30\% of the stars with an error in \te\
smaller than 10\%.

\section{The effect of random and systematic photometric errors}\label{sec:err}
In order to assess the intrinsic precision of the procedure described
in this paper we have tested it on simulations. Artificial stars
were created with $\ebv=0.2$ in the 6 filters listed in  Table~\ref{tab:log},
evenly distributed in the logarithm of effective temperature between 3,500
and 50,000~K, the range covered by \citet{bes98} models. They were,
then, fitted according to the prescriptions of section~\ref{sec:proc}.

\subsection{Random errors}\label{sec:err_rand}
The effects of random errors were mimicked by creating three groups of 10,000
stars and adding to the model magnitudes errors drawn from Gaussians of
different rms (0.02, 0.05 and 0.1~\mg). These stars were then fitted using
only 4, 5 or the complete set of 6 bands.

The situation for a five-band fit (F336W, F439W, F555W, F675W and F814W)
and varying the photometric precision is shown in
Figure~\ref{fig:tout_tin-err}. There we plot the ratio of the output to
input temperature as a function of the input temperature for different
values of the error rms: 0.1~\mg\ in panel~(a), 0.05~\mg\ in panel~(b) and
0.02~\mg\ in panel~(c). In each panel the photometric errors were drawn
from the same Gaussian, independent of the star's temperature. As it can be
seen, there is no systematic trend and the points are mostly scattered around
$\mathrm{T_{out}}=\mathrm{T_{in}}$. The values of the mean and the dispersion
are reported in Table~\ref{tab:rand_err}.

\placefigure{fig:tout_tin-err}

The biggest deviation from the random scatter is represented by the two
features at $\log(\mathrm{T_{in}}) \simeq3.95$, which are most prominent in
panel~(c). These are stars close to the turning point at $\bi{F336W}-\bi{F814W}
\simeq0$ in Figure~\ref{fig:qubit} and were assigned the wrong value of
\ebv\ during the selection process which constitutes the first step of the
procedure. Note that this effect depends on the input photometric error
because the class assignment is based on it.
In the worst case, when the error is 0.02~\mg, however, this 
systematics  affects less than 10\% of the stars at that temperature, split
almost evenly between stars with too high and too low output temperature.
This effect is negligible for studies of stellar populations. However
if one is dealing with special categories of objects, for example variable or
otherwise peculiar stars, which represent a small fraction of the total
population in the $3.92\lesssim\log(\te)\lesssim4$ range one should be aware
that random errors may create false outliers with temperatures deviating
as much as 20\% for up 10\% of the stars in this temperature range.
For this type of studies it is safer not to determine extinction for
individual stars but, rather, to adopt the value of the neighbors not
affected by this problem.

It is interesting to notice in all of the panels in
Figure~\ref{fig:tout_tin-err} that the dispersion increases for hotter stars.
This is because broad band colors become progressively more insensitive to
temperature as it increases. In fact, the scatter at high temperatures becomes
smaller when the F255W filter is added to the fit, as shown in
Figure~\ref{fig:tout_tin-nbands}. Of course, in reality hotter Main Sequence
stars are also brighter than colder ones and, thus, have smaller photometric
error which compensate for the lower sensitivity of colors to high effective
temperatures. Again, see Table~\ref{tab:rand_err} for the values of the mean
and dispersion.

\placefigure{fig:tout_tin-nbands}

\subsection{Systematic errors}\label{sec:err_syst}
\begin{figure}
\efig{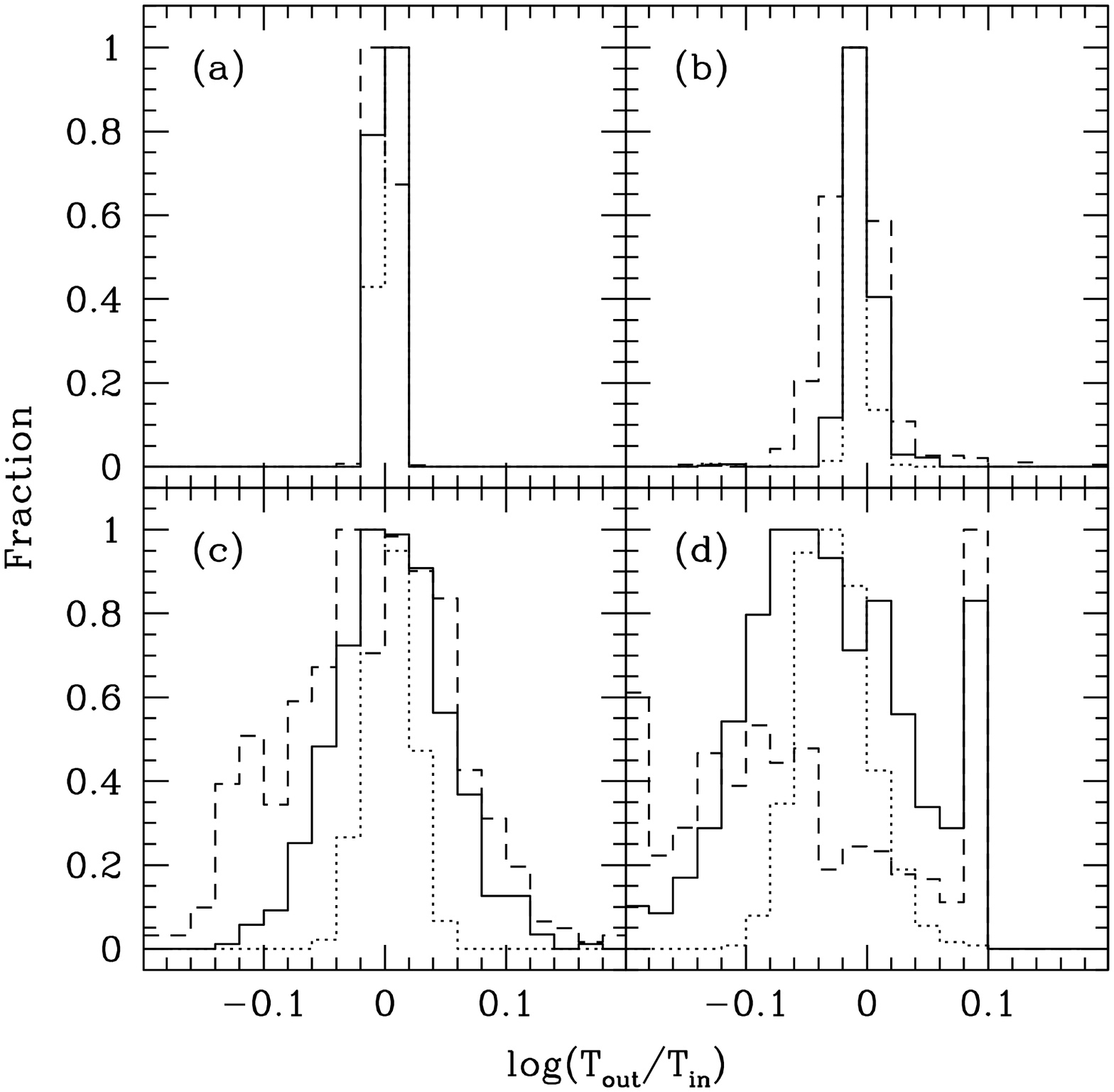}{0.9\linewidth}{0}
\caption{Distribution of the ratio of output to
  input temperature in the case of systematic errors for four representative
  temperatures: 5,000~K in panel~(a), 10,000~K in panel~(b), 20,000~K in
  panel~(c) and 40,000~K in panel~(d). In each panel the dotted line is for
  the case of a systematic error of 0.02~\mg, the full one for 0.05~\mg\ and
  the dashed one for 0.1~\mg.\label{fig:syst_hist}}
\end{figure}

In addition to random errors that vary from star to star and from band
to band the measured fluxes are also affected by systematic errors in
the calibrations of each band, \ie errors in the photometric
zeropoints. These, of course, are different from band to band, but the
same for all the stars in each band. Again, we have explored three
cases (0.02, 0.05 and 0.1~\mg) and the results are shown in
Figure~\ref{fig:syst_hist} that displays the distributions of the
temperature deviations $\log(\mathrm{T_{out}}/\mathrm{T_{in}})$ for four
representative  temperatures, \ie 5,000, 10,000, 20,000 and 40,000~K.  

\placefigure{fig:syst_hist}

From an inspection of Figure~\ref{fig:syst_hist} one notices again that,
given the same error, the temperature measured for hot stars is
intrinsically more uncertain than for cold ones, with the dispersion growing
from less than 1\% for stars at 5000~K, to as much as 12\% at 40,000~K. 
The sharp peak at $\log(\mathrm{T_{out}/T_{in}})\simeq0.1$ found for
$\mathrm{T_{in}}=40,000$~K (Figure~\ref{fig:syst_hist}, panel~(d)) is
populated by stars for which the fit, starting from
$\mathrm{T_{in}}=40,000$~K, would indicate a temperature higher than
50,000~K, the highest in the \citet{bes98} grid. Even though the peak
is just an artifact, it is clear that calibration errors much smaller
than 0.1~\mg\ are necessary for the data to be useful at all
temperatures.

\section{Summary and conclusions}\label{sec:summary}
In this paper we have presented and discussed a new technique that uses
multiband photometry to measure  reddening, effective temperature and, given
the distance, bolometric luminosity for resolved stars. The method is based
on broad band optical and near ultraviolet photometry and, in the test
case presented here for the field around SN1987A in the Large
Magellanic Cloud, yields to a reddening determination every 13~square
arcseconds: a fine grid indeed! Being able to measure the reddening to
thousands of individual stars is of great importance in all the
astrophysical environments in which the interstellar medium is clumpy
and, therefore, produces differential reddening, as is the case, for
example, of virtually every star forming region.

There are 3 unknown quantities in the fit: the radius, the effective
temperature and the reddening. Formally, then, at least three
photometric points are required for each star to make the problem
well defined, and more are needed when dealing with stellar population
encompassing a wide range of temperatures. Obviously, the more bands
are available, the better the result of the fit. As we have discussed
thoroughly, at least one of the bands has to fall  at wavelengths
shorter than the Balmer jump because the optical fluxes alone are not
enough to reliably recover both \te\ and \ebv. On the other hand,
though, in the case in which observing time constraints limit to
one the the number of bands that can be used for UV observations, it is
not necessary nor advisable  to resort to bands at extremely short
wavelengths, and observing just blueward of the Balmer jump, a U-like
band, is the best choice. Of course, observing in more than one
ultraviolet band can only help in reducing the errors on the derived
quantities. Although we have used  the \wfpc\ counterparts to the
classical  Johnson-Cousin passbands to illustrate  our method, it
proved to work almost equally well when broader filters were used.
Broader filters means more photons detected per unit time and,
ultimately, a more efficient use of the telescope time at one's
disposal.

Once the stars are accurately placed in the Hertzprung-Russel diagram,
the physical properties of the stellar population, such as star
formation history, mass function etc, can be studied in detail. In this
respect, let us stress again the importance of determining the
reddening from the same stars which are the ultimate target of the
investigation, without having to resort to uncontrollable assumptions
on the magnitude and distribution of the interstellar extinction. 

To conclude, we refer the reader to \citet{prsk00} for a discussion of
the properties of the young stellar population in the field around
SN1987A and to \citet{rom00} for an application of the dereddend
Color-Magnitude diagram to measuring the distance to the Large
Magellanic Cloud. More work on the derivation of the Initial Mass
Function and the Star Formation Rate in the SN 1987A field is in
progress, and  the results will be presented in a forthcoming paper
\citet{rom02}.

\acknowledgments
It is a pleasure to thank Prof. Giuseppe Bertin for many stimulating
discussions during the development of the method described here.
We gladly acknowledge the comments of an anonymous referee which helped us
greatly to improve the paper both in the contents and the presentation.
MR and SS acknowledge the kind hospitality of STScI during a number of
stays. This work was supported in part by HST-STScI by STScI-DDRF grants
\# 82131, 82160, and 82186 to NP. Support for the SINS program was provided
by NASA through grant GO-9114 from the  Space Telescope Science Institute,
which is operated by the Association of Universities for Research in
Astronomy, Inc., under NASA contract NAS-26555.

\section*{Appendix A: The fitting procedure}\label{sec:app_fit}
As repeatedly stated, the general idea of our method is to recover
simultaneously the extinction along the line of sight and the effective
temperature of a star by fitting the observed magnitudes to the ones computed
using stellar atmosphere models reddened by various amounts of \ebv. In this
Appendix we describe the details of the fitting procedure.

Among the three parameters we want to determine, only two change the
\emph{shape} of the spectrum, \ie \ebv\ and effective temperature (\te). The
third one, \ie the radius $R$, causes just a rigid shift of the
flux on a logarithmic scale. So, we first determine the shape of the spectrum
by solving simultaneously for \ebv\ and \te, then we compute the radius of the
star. For every star, the following steps are performed:

\begin{enumerate}
\item The input spectrum is normalized to the band with the highest photometric
  accuracy, usually the F555W, to eliminate the overall normalization, which
  depends an the star's radius. Obviously, the theoretical
  spectra are normalized in the same way. Let $m_{obs,i}$ be the normalized
  observed magnitude in filter $i$ and $m^0_{ms,ij}$ the normalized model
  magnitude for temperature $j-th$ in the same filter;

\item The theoretical magnitudes are reddened, for various values of \ebv.
  If $\mathcal{R}_i$ is the extinction coefficient in the band $i$ due to the
  LMC dust and $\mathcal{R}_{i,\mathrm{MW}}$ is the one due to our Galaxy:

  \begin{eqnarray}
  m_{ms,ij}\left[\ebv\right]&=m_{ms,ij}^0+\left[\mathcal{R}_i\cdot\ebv\right]+
  \nonumber\\
  &\left[\mathcal{R}_{i,\mathrm{MW}}\cdot\ebv_\mathrm{MW}\right]
  \end{eqnarray}
  with $\ebv_\mathrm{MW}=0.05$, as already discussed. The extinction
  coefficients for selected WFPC2 filters are listed in Table~\ref{tab:abs};

\item For any given temperature and reddening, we have computed the $\chi^2$
  as:

  \begin{equation}
  \chi_j^2\left[\ebv\right]=\sum_{i=1}^{N-1} \left[\frac{m_{ms,ij} \left[\ebv
  \right]-m_{obs,i}}{\delta m_{obs,i}}\right]^2
  \end{equation}
  where $N$ is the number of photometric points and $\delta m_{obs,i}$ is the
  error on the observed magnitude $m_{obs,i}$. The quantity $\delta m_{obs,i}$,
  of course, is the sum in quadrature of the photometric error on the
  magnitude in band $i$ plus the one on the magnitude in the band to which
  the spectrum was normalized. The index $i$ in the sum runs
  on all bands, but the one that was used to normalize the spectrum. 
  If a star is not detected in a certain band, typically in the UV, a
  $3\sigma$ upper limit for the flux is used. The models that predict a flux 
  in that band higher than this limit are rejected.

  Of course, the best-fit model corresponds to the combination temperature,
  and \ebv\ that minimizes $\chi_j^2\left[\ebv\right]$;
\item\label{fa} Once the minimum $\chi_j^2$ is found, the errors in
  temperature ($\delta\te$) and reddening ($\delta\ebv$) are computed from
  the $\chi^2$ map;
\item Now that the shape of the spectrum is determined, we compute the
  \emph{radius} of the star
  as the error-weighted mean of the logarithmic shifts required to match the
  model to the observed flux in every band. Since the theoretical fluxes are
  flux densities at the \emph{surface of the star}, the angular radius $\rho$
  derived from band $i$ is:

  \begin{equation}
  \log\left(\rho_i\right)=-0.2\cdot\left(m_{obs,i}^0-m_{ms,i}^{fit}\right)
  \end{equation}
  where $m_{ms,i}^{fit}$ is the magnitude in filter $i$ of the
  best-fitting theoretical spectrum and $m_{obs,i}^0$ is the observed
  magnitude, dereddened with the the best-fitting value of \ebv. The best
  estimate of the angular radius $\rho$, then, is:

  \begin{equation}
  \rho=\frac{\sum_{i=1}^n \rho_i/\delta^2 m_{obs,i}}
  {\sum_{i=1}^n 1/\delta^2 m_{obs,i}}
  \end{equation}
  where $n$ is the number of bands with actual flux measurements. The
  associated error $\delta\rho$, then, is:

  \begin{equation}
  \left(\delta\rho\right)^2=\frac{\sum_{i=1}^{n}\left(\rho_i-\rho\right)^2/
  \delta^2 m_{obs,i}}{(n-1)\cdot\sum_{i=1}^{n}
  1/\delta^2 m_{obs,i}}
  \end{equation}

  To convert the angular radius of a star to the linear one ($R$), it
  is necessary to know its distance $D$:

  \begin{equation}
  R=\rho\cdot\frac{D}{\mathrm{R}_{\sun}}
  \end{equation} 
  in this case, $R$ is expressed in units of the solar radius
  R$_\sun$.

  Given the uncertainty on the distance, the error on the linear radius
  ($\delta R$) is:

  \begin{equation}
  \frac{\delta R}{R}=\sqrt{\left(\frac{\delta\rho}{\rho}
  \right)^2+\left(\frac{\delta D}{D}\right)^2}
  \end{equation}
  This error does not take into account any intrinsic dispersion in the
  distance of individual stars, which may introduce additional scattering in
  the HR diagram.

  For example, in the case of the LMC, we have adopted the value originally
  measured purely geometrically by \citet{pan91} and later improved by
  \citet{pan99}. Using the rings around SN1987A, they found
  $D=51.4\pm 1.2$~kpc;

\item Finally, the intrinsic luminosity (L) can be computed according to the
  black body law:

  \begin{equation}
  \frac{\mathrm{L}}{\mathrm{L}_\sun}=\left(\frac{R}{\mathrm{R}_\sun}
  \right)^2\cdot \left(\frac{\te}{\mathrm{T}_{eff,\odot}}\right)^4
  \end{equation}
  where $\mathrm{T}_{eff,\odot}$ is the temperature of the Sun.

  Since the errors in temperature and radius are correlated, the error on the
  luminosity cannot be computed in a straightforward way from the previous
  equation. It is more convenient to write the luminosity in another,
  completely equivalent form. Let us first define the quantities:

  \begin{equation}
  l_{obs,i}\equiv10^{-0.4\cdot\left(m_{obs,i}^0-ZP_i\right)}
  \end{equation}
  \ie the observed flux in filter $i$, and:

  \begin{equation}
  b\equiv\frac{\sigma\, \te^4}{\sum_{i=1}^n l_{ms,i}^{fit}}
  \end{equation}
  where $l_{ms,i}^{fit}$ is the flux in filter $i$ of the best-fitting model.
  The quantity $b$ is the equivalent of the bolometric correction, relating
  the total flux in $n$~bands to the bolometric luminosity per unit surface
  $\sigma \te^4$.
  We can now write the luminosity as:

  \begin{equation}  
  L=4\pi D^2\: l_{obs}\: b
  \end{equation}
  where $l_{obs}=\sum_i l_{obs,i}$. All the quantities in this last equation
  are independent from each other and, hence, it is straightforward to compute
  the error on the luminosity:

  \begin{equation}
  \frac{\delta L}{L}=\sqrt{\left(2\,\frac{\delta D}{D}\right)^2+
  \left(\frac{\delta l_{obs}}{l_{obs}}\right)^2+
  \left(\frac{\delta b}{b}\right)^2}
  \end{equation}
  where

  \begin{equation}
  \frac{\delta l_{obs}}{l_{obs}}=\frac{1}{l_{obs}}\sqrt{\sum_{i=1}^n
  \left[\left(10^{0.4\delta m_{obs,i}}-1\right)\cdot l_{i,obs}\right]^2}
  \end{equation}
  and:

  \begin{equation}
  \frac{\delta b}{b}=\frac{db}{d\te} \: \delta\te
  \end{equation}
  In this last equation, it is clear that the first term on the right-hand
  side, $db/d\te$, is completely determined by the models, whereas the second
  one, $\delta\te$, is the error on the temperature from the fit computed in
  step~\ref{fa}.

  In the case of the LMC, $\delta D/D=0.02$ \citep{pan99}.

  Let us stress once again that the stars may not be all at the same distance
  and that this may introduce additional scattering in luminosity. From
  photometry alone, distances can be measured only in a statistical sense for
  an entire population and not for individual stars. However, since the
  intrinsic depth the LMC is estimated to be smaller than 600~pc
  \citep{cro95}, this uncertainty is negligible.
\end{enumerate}

\section*{AppendixB: Broad band colors and extinction coefficients in
selected \wfpc\ filters}\label{sec:app_colors}
The \wfpc\ is equipped with a large set of broad, medium and narrow band
filters for imaging \citep[see][for a complete list and full
details]{wfpcman}. Some of them closely match filters in the standard
Johnson-Cousin (e.g. the F555W and the V, the F439W and the B, etc) or
Str\"omgren (e.g. the F410M and the \emph{v}, the F467M and the \emph{b}, etc)
systems, while some other have no direct counterpart in any standard
ground-based set, \eg the F450W and F606W, just to name two, or, obviously, all
ultraviolet passbands.

Color transformations have been devised to convert the \wfpc\
magnitudes into the corresponding ones in the Johnson-Cousin system,
but, they are limited to only a handful of passbands. Moreover, quoting
directly from \citet{wfpcman}, they ``should be used with caution for
quantitative work'', because they are just first order approximations
and unavoidably degrade the quality of the original photometry. A much
better option, when comparing \wfpc\ photometry to theoretical
expectations, is to compute these latter ones directly in the passbands
used in the observations. As explained in section~\ref{sec:proc},
this is what we have done in this paper, thus minimizing the
uncertainties in the comparison between the observed \wfpc\ magnitudes
and the synthetic ones computed using the \citet{bes98} models.

In Tables~\ref{tab:cas22} and~\ref{tab:cas63} we list the magnitudes in
widely used \wfpc\ passbands for solar metallicity and for
[Fe/H]$=-0.5$, the appropriate value for the LMC, respectively. They
were computed using the \emph{IRAF-synphot} synthetic photometry
package and, for the sake of generality, the magnitudes in the
ultraviolet filters are referred to the decontaminated camera.
The throughput degradation due to the accumulation of material between
decontamination can be easily computed with the formul\ae\ given  in
\citet{bagg96}. The magnitudes are for a 1~R$_\sun$ star as seen from a
distance of 10~pc and scaling them for any radius and distance using
equation~(\ref{eq:redmag}) is straightforward.

The other important quantities that one needs in order to interpret the
data are the extinction coefficients. Again, it is very important to
compute them in the very same passbands used for the observations. The
values of the normalized extinction $\mathcal{R}_\lambda =
A_\lambda/E(B-V)$ for the filters of Tables~\ref{tab:cas22}
and~\ref{tab:cas63} computed for the galactic reddening law as
compiled by \citet{savmat79} and the models atmospheres by
\citet{bes98} are listed in Table~\ref{tab:abs} for various values of
the effective temperature and the color excess \ebv.

Using the entries of Tables~\ref{tab:cas22}, \ref{tab:cas63} and
\ref{tab:abs}, then, it is very easy to model a star of metallicity Z,
surface gravity $g$, effective temperature \te, radius $R$ (in solar units)
as seen at a distance $D$ (in parsecs) through a layer of dust characterized
by a reddening \ebv:

\begin{eqnarray}
m(\mathrm{Z},g,\te,R,D,\ebv)=m_0(\mathrm{Z},g,\te)+\nonumber\\
 -5\cdot\left(\frac{R}{1~\mathrm{R}_\sun}\right)
+5\cdot\log\left(\frac{D}{10~pc}\right)\nonumber \\
 +\mathcal{R}(\te,\ebv)\cdot\ebv
\label{eq:redmag}
\end{eqnarray}

\noindent where $m_0(\mathrm{Z},g,\te,Z)$ is the appropriate value,
either read or interpolated  from Table~\ref{tab:cas22} or
\ref{tab:cas63}, and $\mathcal{R}(\te,\ebv)$ can be taken from
Table~\ref{tab:abs}.

Ultraviolet \wfpc\ filters are notoriously affected by red leak, and a
quite substantial one when red stars are observed \citep[see table 3.13
of][]{wfpcman}. This is the reason why, for example, as shown in
Table~\ref{tab:abs}, the extinction coefficient $\mathcal{R}$ in the
F170W filter is about 4 times smaller for a $\te=3,500$~K star than
for a 40,000~K one! One can easily assess the influence of the red leak
by recomputing the same coefficients after cutting it out from the
response curve of the filter, in this case the F170W. What one finds,
then, is that, while the extinction coefficient for $\te=40,000$~K is
virtually unchanged, the one for $\te=3,500$~K increases by a factor
of 3, thus reducing the difference between them to about 30\%. The
residual difference in $\mathcal{R}$ between different temperatures,
the lower one yielding to lower extinction coefficients, is caused by
the fact that filters have a finite, non negligible bandwidth and,
thus, the effective extinction in a band depends on the shape of the
spectrum which is a function of temperature. The same explanation
applies for the fact (also illustrated in Table~\ref{tab:abs}) that
extinction coefficients become smaller as the optical depth of the
dust increases because dust extinction makes a spectral energy
distribution become redder.

Tables with similar quantities computed for all of WFPC2 filters and
for all model atmospheres in the \citet{bes98} grid are available and
may be provided on request.

\clearpage
\clearpage
\vspace*{3cm}\centerline{\texttt{See 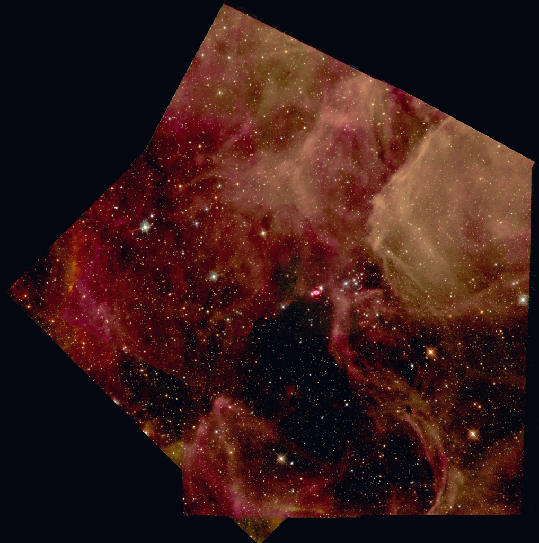}}\vspace{0.3cm}
\figcaption{The field centered on SN1987A (about 130\pp\
  radius) as observed in the combination of the B, V, and I broad bands plus
  the [OIII] and H$\alpha$ narrow band images.\label{fig:truecolor}}

\vspace{5cm}\centerline{\texttt{See 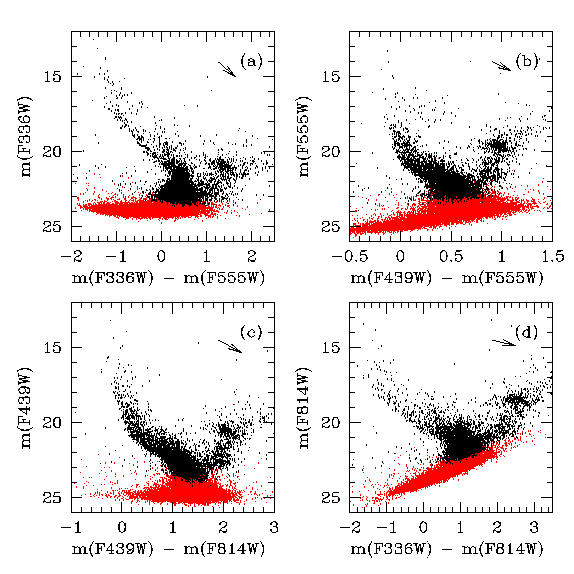}}\vspace{0.3cm}
\figcaption{Color-Magnitude Diagrams for four combination of
  filters. The grey dots are stars with average errors, as defined in
  equation~(\ref{eq:mer}), $\dfive>0.1$, whereas the black dots
  are the 6,695 stars with $\dfive<0.1$. The reddening vector corresponding
  to $\ebv=0.2$ is show in each panel\label{fig:cmd}}

\vspace{5cm}\centerline{\texttt{See 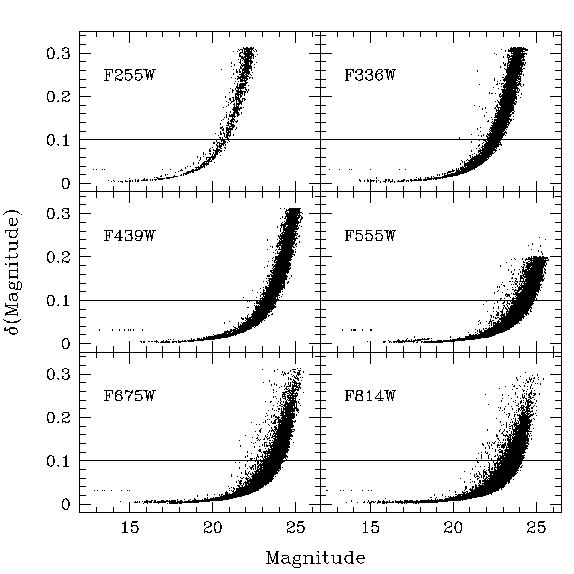}}\vspace{0.3cm}
\figcaption{Photometric error as a function of magnitude for
  the 6~broad band filters around SN1987A. The 0.1~\mg\ error level is
  indicated.\label{fig:phot_err}}

\clearpage

\clearpage

\clearpage

\clearpage

\clearpage

\vspace{3cm}\centerline{\texttt{See 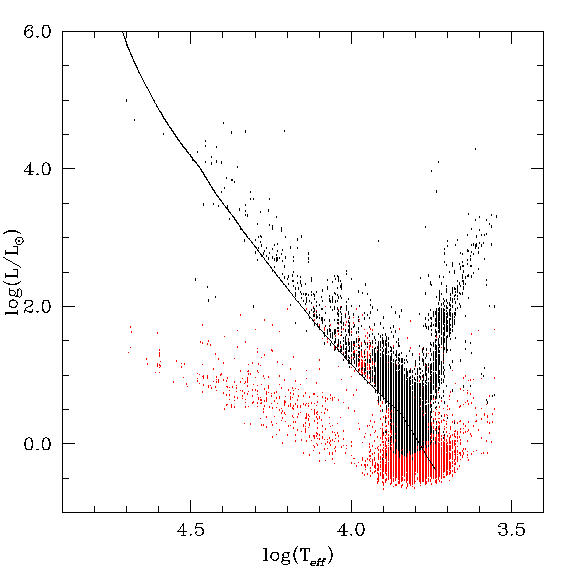}}\vspace{0.3cm}
\figcaption{HR diagram for the stars in the field around
  SN1987A. The black dots are the 9,474 stars with $\delta\log(\te)<0.05$.
  The Zero Age Main Sequence for Z$=0.3\cdot\mathrm{Z}_\sun$ from the models
  by \protect{\citet{bc93}} and \protect{\citet{ccs94}} for masses below
  25~M$_\sun$ and by \citet{sch93} above 25~M$_\sun$ is shown as a full line.
  \label{fig:sn87a_hrms}}

\clearpage

\vspace{5cm}\centerline{\texttt{See 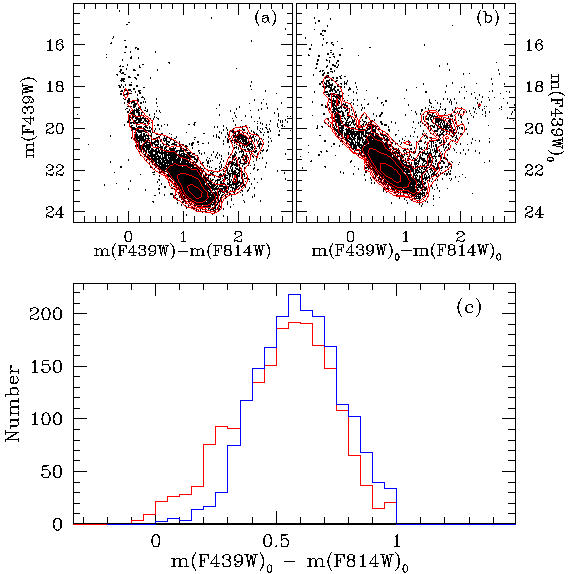}}\vspace{0.3cm}
\figcaption{Before and after the reddening correction.
  The observed \bi{F439W}~vs.~$\bi{F439W}-\bi{F814W}$ diagram for all the stars
  in our \wfpc\
  frames with $\dfive<0.1$ is shown in panel~(a), while the
  dereddened one in panel~(b). Black dots represent the stars for which \ebv\
  was determined individually (class I and II, see text), grey dots stars
  dereddened with the mean value of \ebv\ from the closest class~I and~II
  neighbors. The density contours, spaced by  factors of 2, are are superposed
  in red. The visual impression that the Main Sequence in panel~(b) is narrower
  than the one in panel~(a) is confirmed by the histograms shown in panel~(c).
  Here, the blue line represents the color distribution of the dereddened
  stars in the $21<\mathrm{m(F439W)_0}<22$ magnitude interval and the red one
  the distribution of the observed colors for the same stars. For the
  sake of clarity, the latter was shifted by 0.5~\mg\ in color to have the
  same mode of the dereddened one. There are 218 and 192 stars, respectively,
  at the peak of the dereddened and observed distribution, while both
  histograms contain a total of 1923 stars.\label{fig:beaft_bihall}}

\vspace{1cm}

\clearpage
\vspace{2cm}\centerline{\texttt{See 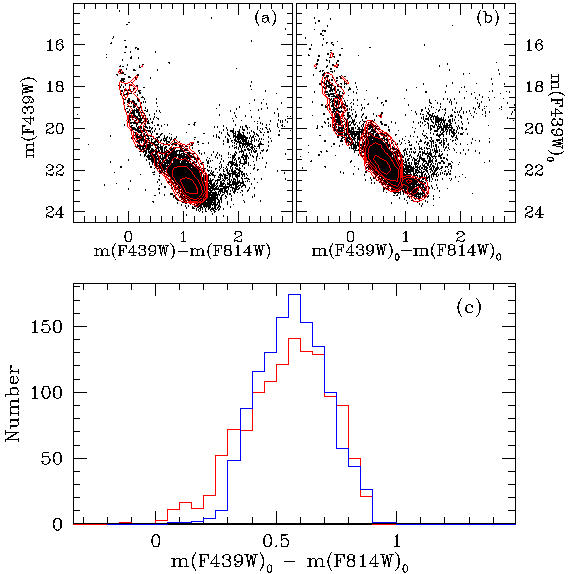}}\vspace{0.3cm}
\figcaption{Same as Figure~\ref{fig:beaft_bihall}, but
  the contours in panels~(a) and (b) and the histograms in panel~(c) refer only
  to the stars with individual reddening correction (class~I and II, see
  text). The peak values of the histograms in panel~(c) are 174 and 141
  for the corrected (blue line) and observed one (red line),
  respectively.\label{fig:beaft_bihind}}

\vspace{2cm}

\clearpage

\clearpage

\vspace{2cm}
\hspace{-0.25cm}\epsfig{file=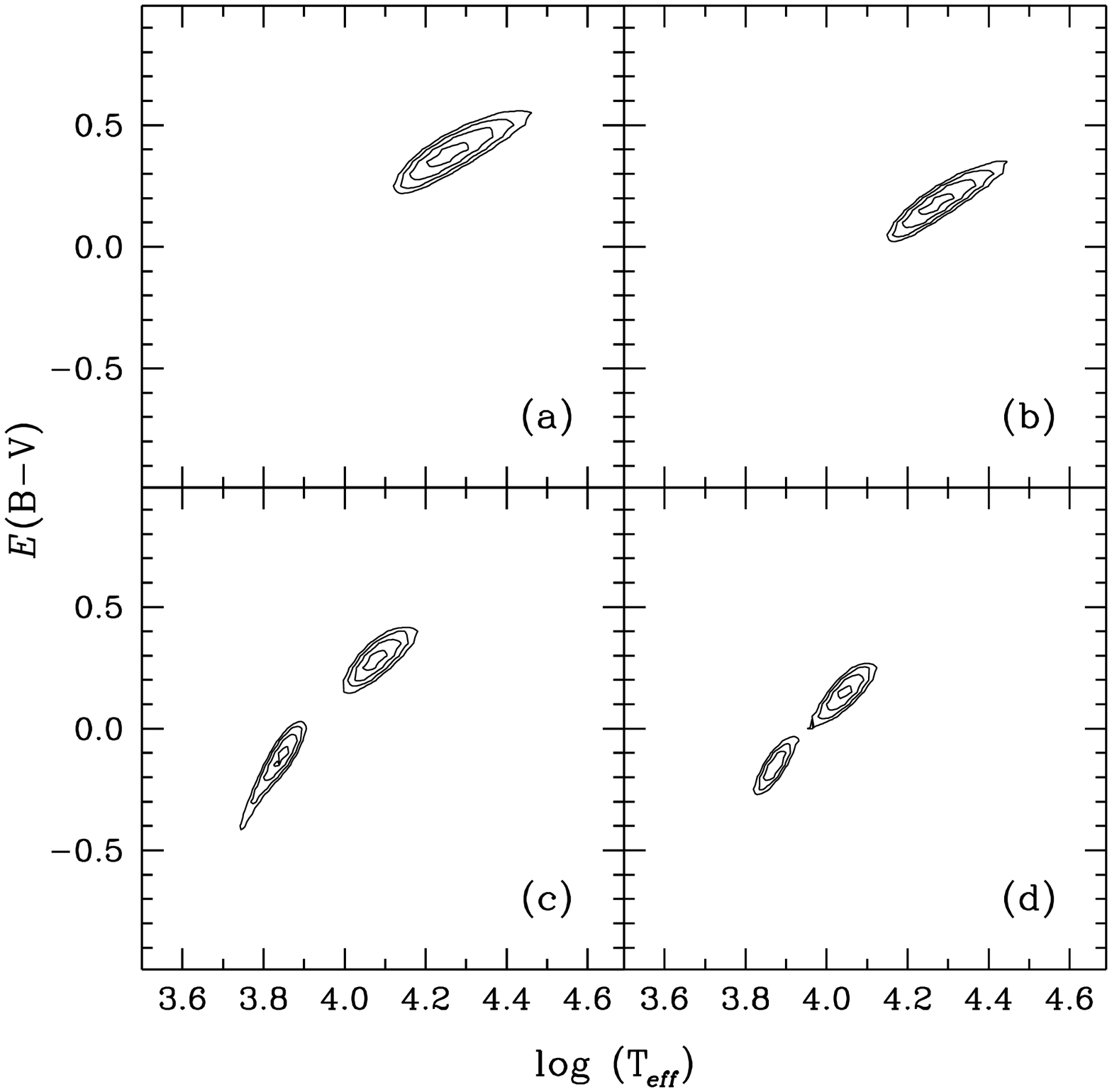, height=0.49\linewidth}\hfill\texttt{See 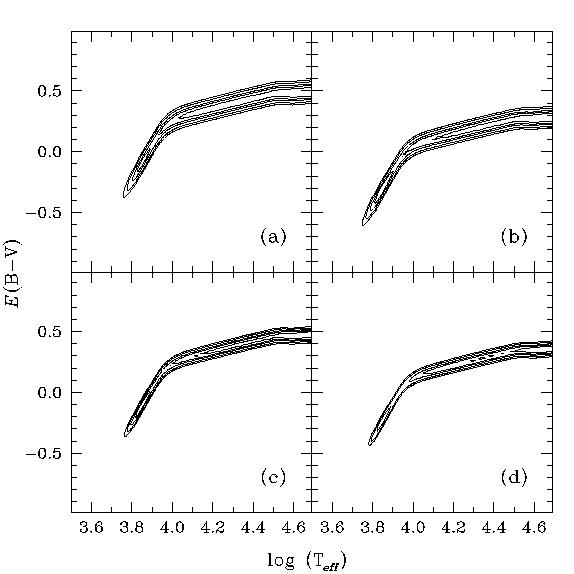}\hspace{3cm}
\figcaption{Contour plots in the \te-\ebv\
  space for four stars with unambiguous solution. \emph{Left panel:} only four
  bands (F439W, F555W, F675W and F814W) are used in the fit. \emph{Right
  panel:} all six bands are used. In both panels the contours are for
  $\delta\chi^2=2$ from the minimum.\label{fig:mchi_I}}

\clearpage

\clearpage

\clearpage
\vspace*{3cm}\centerline{\texttt{See 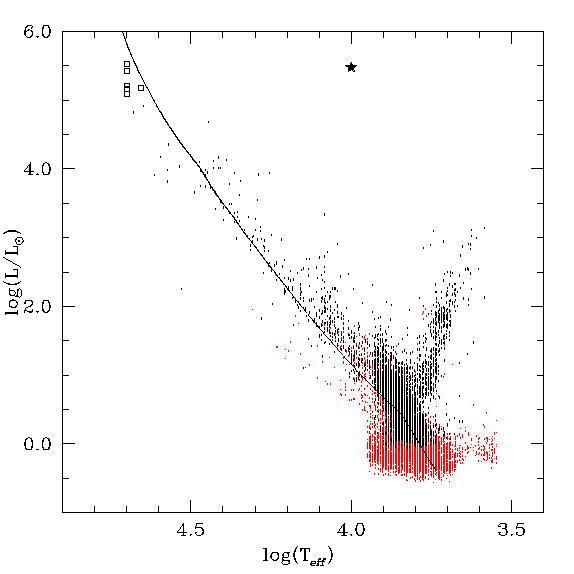}}\vspace{0.3cm}
\figcaption{HR diagram for the 13,098 stars detected in
  the control field for SN1987A. Black dots are the 4,912 stars with
  $\delta\log(\te)<0.05$. The open squares indicate stars for which the F300W
  magnitude is ill determined because of saturation and, hence, the fit was
  performed excluding this filter. The location of \mbox{Sk~-69~211} according
  to the photometry by \protect{\citet{fitz88}} is shown with a star symbol.
  The Zero Age Main Sequence for Z$=0.3\cdot\mathrm{Z}_\sun$ from the models
  by \protect{\citet{bc93}} and \protect{\citet{ccs94}} for masses below
  25~M$_\sun$ and by \citet{sch93} above 25~M$_\sun$ is shown as a full line.
  \label{fig:sn87a_parf_hr}}

\vspace{2cm}\centerline{\texttt{See 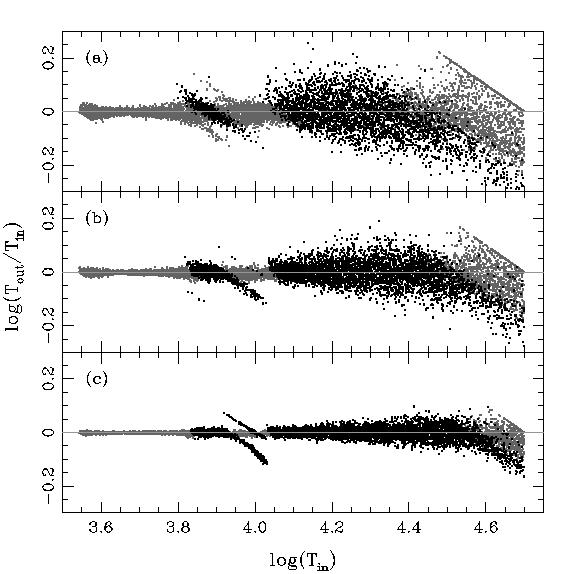}}\vspace{0.3cm}
\figcaption{Ratio of output to input temperature as
  a function of the latter for 10,000 model stars distributed evenly
  in $\log(\mathrm{T_{in}})$. The fit was performed as described in
  section~\protect{\ref{sec:proc}} using 5 bands: F336W, F439W, F555W,
  F675W and F814W. The input stars in panel~(a) simulate a random photometric
  error of 0.1~\mg\ in all of the bands, those in panel~(b) an error of
  0.05~\mg\ and those in panel~(c) an error of 0.02~mg. Black dots are
  class~I and II stars for which the reddening was determined individually,
  while grey ones are class~III stars dereddened with the average value
  of their class~I and II neighbors.\label{fig:tout_tin-err}}

\vspace{2cm}\centerline{\texttt{See 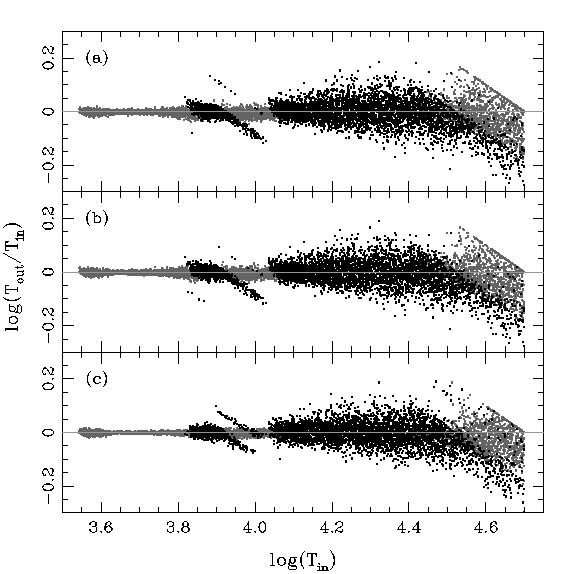}}\vspace{0.3cm}
\figcaption{Same as
  Figure~\protect{\ref{fig:tout_tin-err}}, but the fit was performed with
  different combinations of bands: 4 in panel~(a) (F336W, F439W, F555W and
  F814W), 5 in panel~(b) (those of panel~(a) plus F675W) and 6 in panel~(c)
  (those of panel~(b) plus F255W). The stars simulate a random photometric
  error of 0.05~\mg.\label{fig:tout_tin-nbands}}

\clearpage

\clearpage
\begin{table*}[!ht]
\begin{center}
\caption{Log of the observations centered on Supernova~1987A.}\label{tab:log}
\begin{tabular}{*{5}{c}}
& & & & \\
Filter Name &  \multicolumn{3}{c}{Exposure Time (seconds)}  & Comments\\
\cline{2-4} &September 1994\tablenotemark{a}
&February 1996\tablenotemark{b}
&July 1997\tablenotemark{c}   & \\ \tableline
{\bf F255W} &      2x900      &    1100+1400   &   2x1300    & UV~Filter  \\
{\bf F336W} &      2x600      &      2x600     &   2x800     & ~U~Filter  \\
{\bf F439W} &      2x400      &     350+600    &   2x400     & ~B~Filter  \\
{\bf F555W} &      2x300      &      2x300     &   2x300     & ~V~Filter  \\
{\bf F675W} &      2x300      &      2x300     &   2x300     & ~R~Filter  \\
{\bf F814W} &      2x300      &      2x300     &   2x400     & ~I~Filter
\end{tabular}
\end{center}
\tablenotetext{a}{September 24, 1994, proposal number 5753.}
\tablenotetext{b}{February 6, 1996, proposal number 6020.}
\tablenotetext{c}{July 10, 1997, proposal number 6437.}
\end{table*}

\begin{table}[!ht]
\begin{center}
\caption{Log of the observations of the control field for SN1987A. They were
taken on January, 1$^\mathrm{st}$ 1997 and the proposal number is
6437.}\label{tab:sn87a_parf_log}
\begin{tabular}[b]{*{3}{c}}
Filter Name & Exposure Time (s) &     Comments   \\ \hline
{\bf F300W} &    $600+1200$     &  Wide U filter \\
{\bf F450W} &     $50+200$      &  Wide B filter \\
{\bf F675W} &     $50+200$      &     R filter   \\
{\bf F814W} &     $140+400$     &     I filter
\end{tabular}
\end{center}
\end{table}

\begin{table}[!ht]
\begin{center}
\caption{Observed magnitudes and statistical errors for the stars of
Figure~\protect{\ref{fig:wbsp_nfit}}.}\label{tab:wbsp_nfit}
\begin{tabular}[b]{*{7}{c}}
Panel & F255W & F336W & F439W & F555W & F675W & F814W\\\tableline
 (a) & $15.74\pm0.01$ & $16.13\pm0.01$ & $17.37\pm0.01$ & $17.23\pm0.01$ & $17.44\pm0.01$ & $17.48\pm0.01$ \\ 
 (b) & $19.34\pm0.04$ & $19.24\pm0.01$ & $19.83\pm0.01$ & $19.69\pm0.01$ & $19.45\pm0.01$ & $19.28\pm0.01$ \\ 
 (c) & $21.50\pm0.20$ & $20.99\pm0.03$ & $20.91\pm0.02$ & $20.71\pm0.01$ & $20.45\pm0.01$ & $20.35\pm0.01$ \\ 
 (d) & $22.10\pm0.30$ & $21.32\pm0.04$ & $21.23\pm0.02$ & $20.85\pm0.01$ & $20.46\pm0.01$ & $20.26\pm0.01$ \\ 
\end{tabular}
\end{center}
\end{table}

\begin{table}[!ht]
\begin{center}
\caption{Best-fit parameters for the spectra of
Figure~\protect{\ref{fig:wbsp_fit}}. The star in panel~(a) is saturated and
this causes the large quoted errors on the quantities derived for it.
The corresponding magnitudes are listed in Table~\protect{\ref{tab:wbsp_nfit}}.
\label{tab:wbsp_fit}}
\begin{tabular}{ccccc}
Panel &     \te (K)     &     \ebv        & log(L/L$_\sun$)  & R/R$_\sun$     \\\tableline
  (a) &$25,350\pm5,700$ & $0.202\pm0.080$ & $ 3.662\pm0.264$ &$3.444\pm0.090$ \\
  (b) &$18,160\pm2,100$ & $0.431\pm0.051$ & $ 2.695\pm0.166$ &$2.265\pm0.050$ \\
  (c) &$ 8,000\pm 350$  & $0.117\pm0.050$ & $ 1.176\pm0.079$ &$2.070\pm0.042$ \\
  (d) &$ 7,480\pm 300$  & $0.209\pm0.047$ & $ 1.255\pm0.067$ &$2.548\pm0.052$ \\
\end{tabular}
\end{center}
\end{table}

\begin{table}[!ht]
\begin{center}
\caption{Mean ratio and rms scatter of the output to input values of selected
quantities for 10,000 artificial stars (see text).}\label{tab:rand_err}
\begin{tabular}{*8{c}}
Input & Fitted & \multicolumn{2}{c}{\te} & \multicolumn{2}{c}{\ebv} & \multicolumn{2}{c}{L} \\\cline{3-8}
error & bands & $\mathrm{T_{in}}<9,000$ & $\mathrm{T_{in}}>9,000$
& $\mathrm{T_{in}}<9,000$ & $\mathrm{T_{in}}>9,000$
& $\mathrm{T_{in}}<9,000$ & $\mathrm{T_{in}}>9,000$\\
(rms) & & & &\\\tableline
0.1 & 5 & $0.99\pm0.05$ & $0.97\pm0.18$ & $0.01\pm0.02$ & $-0.02\pm0.06$ & $0.98\pm0.17$ & $1.01\pm0.58$ \\
0.05& 5 & $0.99\pm0.03$ & $0.98\pm0.12$ & $0.01\pm0.02$ & $-0.01\pm0.04$ & $0.97\pm0.10$ & $0.99\pm0.37$ \\
0.02& 5 & $0.99\pm0.04$ & $0.99\pm0.06$ & $0.01\pm0.03$ & $-0.00\pm0.02$ & $0.97\pm0.11$ & $0.97\pm0.18$ \\
0.05& 4 & $0.99\pm0.04$ & $0.98\pm0.12$ & $0.01\pm0.02$ & $-0.01\pm0.04$ & $0.98\pm0.11$ & $0.99\pm0.38$ \\
0.05& 6 & $1.00\pm0.03$ & $0.98\pm0.11$ & $0.01\pm0.02$ & $-0.01\pm0.04$ & $0.99\pm0.10$ & $0.97\pm0.35$ \\
\end{tabular}
\end{center}
\end{table}

\begin{deluxetable}{ccccccccccc}
\tabletypesize{\footnotesize}
\tablecolumns{11}
\tablecaption{Magnitudes in selected \wfpc\ broad band filters from the
\protect{\citet{bes98}} models for Z=Z$_\sun=0.02$ referred to a 1~R$_\sun$
star at a distance of 10~pc (see equation~(\ref{eq:redmag})).\label{tab:cas22}}

\tablehead{
\te  &  F170W  &  F255W  &  F300W  &  F336W  &  F439W  &  F450W & F555W  &  F606W & F675W  &   F814W}
\startdata

\\
{\boldmath ${\log(g)=3}$}\\
   3500 &  11.930 &  16.528 &  12.843 &  12.115 &  10.320 &  10.094 &   8.969 &   8.566 &   8.115 &   7.003 \\ 
   4000 &  10.692 &  14.882 &  11.338 &  10.494 &   8.817 &   8.407 &   7.308 &   6.864 &   6.399 &   5.835 \\ 
   4500 &   9.900 &  12.670 &   9.480 &   8.604 &   7.604 &   7.189 &   6.334 &   5.968 &   5.570 &   5.160 \\ 
   5000 &   9.250 &   9.935 &   7.882 &   7.172 &   6.652 &   6.305 &   5.614 &   5.319 &   4.990 &   4.662 \\ 
   5500 &   8.665 &   8.278 &   6.710 &   6.122 &   5.860 &   5.600 &   5.051 &   4.809 &   4.533 &   4.265 \\ 
   6000 &   8.085 &   7.105 &   5.850 &   5.351 &   5.196 &   5.008 &   4.585 &   4.390 &   4.162 &   3.945 \\ 
   6500 &   7.381 &   6.183 &   5.207 &   4.793 &   4.629 &   4.497 &   4.180 &   4.029 &   3.847 &   3.678 \\ 
   7000 &   6.501 &   5.442 &   4.712 &   4.381 &   4.142 &   4.051 &   3.820 &   3.709 &   3.573 &   3.449 \\ 
   7500 &   5.647 &   4.829 &   4.295 &   4.040 &   3.707 &   3.652 &   3.502 &   3.430 &   3.340 &   3.255 \\ 
   8000 &   4.836 &   4.274 &   3.898 &   3.711 &   3.338 &   3.312 &   3.234 &   3.195 &   3.144 &   3.091 \\ 
   8500 &   4.077 &   3.745 &   3.498 &   3.375 &   3.057 &   3.045 &   3.018 &   3.000 &   2.977 &   2.947 \\ 
   9000 &   3.415 &   3.278 &   3.128 &   3.058 &   2.844 &   2.840 &   2.849 &   2.844 &   2.837 &   2.824 \\ 
   9500 &   2.886 &   2.869 &   2.790 &   2.762 &   2.675 &   2.674 &   2.706 &   2.709 &   2.713 &   2.712 \\ 
  10000 &   2.456 &   2.503 &   2.474 &   2.479 &   2.532 &   2.532 &   2.580 &   2.587 &   2.598 &   2.608 \\ 
  10500 &   2.089 &   2.178 &   2.184 &   2.212 &   2.405 &   2.406 &   2.465 &   2.476 &   2.491 &   2.509 \\ 
  11000 &   1.768 &   1.886 &   1.919 &   1.965 &   2.291 &   2.293 &   2.360 &   2.373 &   2.391 &   2.418 \\ 
  11500 &   1.480 &   1.622 &   1.680 &   1.742 &   2.186 &   2.189 &   2.264 &   2.279 &   2.300 &   2.333 \\ 
  12000 &   1.220 &   1.381 &   1.464 &   1.544 &   2.090 &   2.093 &   2.175 &   2.192 &   2.217 &   2.256 \\ 
  12500 &   0.981 &   1.161 &   1.268 &   1.367 &   2.000 &   2.004 &   2.093 &   2.112 &   2.140 &   2.185 \\ 
  13000 &   0.762 &   0.959 &   1.091 &   1.208 &   1.916 &   1.921 &   2.016 &   2.038 &   2.069 &   2.119 \\ 
  15000 &   0.034 &   0.290 &   0.505 &   0.685 &   1.622 &   1.630 &   1.747 &   1.778 &   1.821 &   1.889 \\ 
  17000 &  -0.537 &  -0.242 &   0.034 &   0.263 &   1.367 &   1.376 &   1.513 &   1.551 &   1.606 &   1.690 \\ 
  19000 &  -1.016 &  -0.695 &  -0.374 &  -0.108 &   1.135 &   1.145 &   1.296 &   1.341 &   1.406 &   1.503 \\ 
  21000 &  -1.441 &  -1.091 &  -0.737 &  -0.445 &   0.921 &   0.930 &   1.091 &   1.141 &   1.212 &   1.320 \\ 
  23000 &  -1.832 &  -1.435 &  -1.053 &  -0.741 &   0.724 &   0.733 &   0.904 &   0.956 &   1.032 &   1.147 \\ 
  25000 &  -2.197 &  -1.737 &  -1.337 &  -1.008 &   0.537 &   0.546 &   0.726 &   0.782 &   0.863 &   0.985 \\ 
  27000 &  -2.492 &  -1.918 &  -1.494 &  -1.143 &   0.395 &   0.410 &   0.605 &   0.670 &   0.763 &   0.896 \\ 

\\
{\boldmath ${\log(g)=4.5}$}\\
   3500 &  11.574 &  15.806 &  12.299 &  11.493 &  10.049 &   9.738 &   8.511 &   8.041 &   7.543 &   6.645 \\ 
   4000 &  10.653 &  14.317 &  11.056 &  10.202 &   8.763 &   8.418 &   7.303 &   6.847 &   6.373 &   5.793 \\ 
   4500 &   9.904 &  12.589 &   9.514 &   8.630 &   7.584 &   7.239 &   6.357 &   5.979 &   5.567 &   5.155 \\ 
   5000 &   9.250 &   9.912 &   7.857 &   7.118 &   6.630 &   6.310 &   5.624 &   5.326 &   4.991 &   4.662 \\ 
   5500 &   8.654 &   8.115 &   6.609 &   6.013 &   5.875 &   5.614 &   5.060 &   4.816 &   4.537 &   4.267 \\ 
   6000 &   8.047 &   6.844 &   5.666 &   5.170 &   5.244 &   5.046 &   4.605 &   4.405 &   4.170 &   3.947 \\ 
   6500 &   7.301 &   5.855 &   4.947 &   4.538 &   4.712 &   4.565 &   4.220 &   4.057 &   3.863 &   3.681 \\ 
   7000 &   6.337 &   5.065 &   4.399 &   4.074 &   4.256 &   4.149 &   3.880 &   3.751 &   3.595 &   3.453 \\ 
   7500 &   5.459 &   4.433 &   3.971 &   3.723 &   3.857 &   3.779 &   3.569 &   3.473 &   3.355 &   3.251 \\ 
   8000 &   4.703 &   3.913 &   3.607 &   3.429 &   3.502 &   3.447 &   3.289 &   3.223 &   3.142 &   3.074 \\ 
   8500 &   3.980 &   3.472 &   3.283 &   3.161 &   3.187 &   3.156 &   3.050 &   3.013 &   2.966 &   2.928 \\ 
   9000 &   3.330 &   3.082 &   2.979 &   2.904 &   2.935 &   2.922 &   2.860 &   2.844 &   2.823 &   2.807 \\ 
   9500 &   2.801 &   2.731 &   2.694 &   2.661 &   2.740 &   2.737 &   2.708 &   2.705 &   2.701 &   2.703 \\ 
  10000 &   2.364 &   2.417 &   2.431 &   2.434 &   2.580 &   2.583 &   2.581 &   2.587 &   2.595 &   2.609 \\ 
  10500 &   1.999 &   2.128 &   2.182 &   2.215 &   2.443 &   2.449 &   2.468 &   2.480 &   2.497 &   2.521 \\ 
  11000 &   1.682 &   1.863 &   1.948 &   2.006 &   2.322 &   2.330 &   2.366 &   2.383 &   2.406 &   2.437 \\ 
  11500 &   1.401 &   1.620 &   1.728 &   1.806 &   2.213 &   2.223 &   2.272 &   2.292 &   2.319 &   2.358 \\ 
  12000 &   1.146 &   1.396 &   1.524 &   1.618 &   2.114 &   2.124 &   2.183 &   2.206 &   2.237 &   2.282 \\ 
  12500 &   0.914 &   1.190 &   1.334 &   1.443 &   2.022 &   2.033 &   2.101 &   2.126 &   2.160 &   2.211 \\ 
  13000 &   0.700 &   0.999 &   1.160 &   1.282 &   1.937 &   1.949 &   2.024 &   2.051 &   2.088 &   2.144 \\ 
  15000 &  -0.016 &   0.361 &   0.582 &   0.756 &   1.644 &   1.658 &   1.758 &   1.793 &   1.841 &   1.913 \\ 
  17000 &  -0.577 &  -0.137 &   0.137 &   0.358 &   1.401 &   1.418 &   1.539 &   1.581 &   1.639 &   1.724 \\ 
  19000 &  -1.040 &  -0.550 &  -0.232 &   0.025 &   1.188 &   1.206 &   1.345 &   1.393 &   1.461 &   1.557 \\ 
  21000 &  -1.440 &  -0.909 &  -0.557 &  -0.270 &   0.991 &   1.011 &   1.165 &   1.220 &   1.295 &   1.403 \\ 
  23000 &  -1.800 &  -1.231 &  -0.851 &  -0.540 &   0.807 &   0.827 &   0.995 &   1.054 &   1.137 &   1.254 \\ 
  25000 &  -2.132 &  -1.525 &  -1.121 &  -0.789 &   0.633 &   0.653 &   0.831 &   0.895 &   0.984 &   1.109 \\ 
  27000 &  -2.446 &  -1.792 &  -1.367 &  -1.018 &   0.468 &   0.488 &   0.676 &   0.743 &   0.839 &   0.971 \\ 
  29000 &  -2.744 &  -2.040 &  -1.597 &  -1.231 &   0.307 &   0.328 &   0.527 &   0.598 &   0.699 &   0.838 \\ 
  31000 &  -3.015 &  -2.271 &  -1.816 &  -1.437 &   0.147 &   0.169 &   0.378 &   0.452 &   0.558 &   0.704 \\ 
  33000 &  -3.238 &  -2.472 &  -2.012 &  -1.625 &  -0.002 &   0.020 &   0.237 &   0.313 &   0.423 &   0.572 \\ 
  35000 &  -3.412 &  -2.632 &  -2.170 &  -1.781 &  -0.127 &  -0.104 &   0.117 &   0.193 &   0.304 &   0.456 \\ 
  37000 &  -3.551 &  -2.755 &  -2.292 &  -1.901 &  -0.226 &  -0.204 &   0.021 &   0.097 &   0.207 &   0.360 \\ 
  39000 &  -3.666 &  -2.851 &  -2.386 &  -1.993 &  -0.304 &  -0.281 &  -0.053 &   0.023 &   0.132 &   0.286 \\ 
  41000 &  -3.768 &  -2.933 &  -2.465 &  -2.069 &  -0.370 &  -0.347 &  -0.115 &  -0.039 &   0.071 &   0.226 \\ 
  43000 &  -3.862 &  -3.008 &  -2.537 &  -2.139 &  -0.431 &  -0.407 &  -0.173 &  -0.096 &   0.015 &   0.171 \\ 
  45000 &  -3.949 &  -3.077 &  -2.605 &  -2.204 &  -0.489 &  -0.465 &  -0.228 &  -0.150 &  -0.039 &   0.118 \\ 
  47000 &  -4.030 &  -3.142 &  -2.668 &  -2.266 &  -0.543 &  -0.519 &  -0.281 &  -0.202 &  -0.090 &   0.067 \\ 
  49000 &  -4.106 &  -3.202 &  -2.727 &  -2.323 &  -0.595 &  -0.570 &  -0.331 &  -0.251 &  -0.139 &   0.019 \\ 
\enddata
\end{deluxetable}

\begin{deluxetable}{ccccccccccc}
\tabletypesize{\footnotesize}
\tablecolumns{11}
\tablecaption{Magnitudes in selected \wfpc\ broad band filters from the
\protect{\citet{bes98}} models for Z$=$0.3\,Z$_\sun=6\cdot10^{-3}$ referred
to a 1~R$_\sun$ star at a distance of 10~pc (see equation~(\ref{eq:redmag})).
\label{tab:cas63}}

\tablehead{
\te  &  F170W  &  F255W  &  F300W  &  F336W  &  F439W  &  F450W & F555W  &  F606W & F675W  &   F814W}
\startdata

\\
{\boldmath ${\log(g)=3}$}\\
   3500 &  11.592 &  16.214 &  12.581 &  11.833 &  10.214 &   9.855 &   8.598 &   8.116 &   7.605 &   6.648 \\ 
   4000 &  10.663 &  14.631 &  11.027 &  10.147 &   8.732 &   8.331 &   7.282 &   6.832 &   6.354 &   5.809 \\ 
   4500 &   9.877 &  11.619 &   9.048 &   8.245 &   7.504 &   7.120 &   6.323 &   5.973 &   5.585 &   5.166 \\ 
   5000 &   9.212 &   9.214 &   7.539 &   6.909 &   6.576 &   6.261 &   5.621 &   5.338 &   5.016 &   4.675 \\ 
   5500 &   8.586 &   7.704 &   6.430 &   5.910 &   5.811 &   5.576 &   5.074 &   4.841 &   4.570 &   4.292 \\ 
   6000 &   7.880 &   6.595 &   5.617 &   5.189 &   5.168 &   5.001 &   4.620 &   4.431 &   4.206 &   3.980 \\ 
   6500 &   7.004 &   5.761 &   5.035 &   4.694 &   4.622 &   4.504 &   4.218 &   4.071 &   3.891 &   3.713 \\ 
   7000 &   6.157 &   5.119 &   4.591 &   4.329 &   4.147 &   4.064 &   3.856 &   3.747 &   3.613 &   3.480 \\ 
   7500 &   5.395 &   4.590 &   4.211 &   4.017 &   3.722 &   3.670 &   3.533 &   3.463 &   3.373 &   3.281 \\ 
   8000 &   4.637 &   4.114 &   3.848 &   3.711 &   3.354 &   3.331 &   3.263 &   3.225 &   3.174 &   3.114 \\ 
   8500 &   3.927 &   3.646 &   3.478 &   3.394 &   3.078 &   3.068 &   3.045 &   3.028 &   3.004 &   2.969 \\ 
   9000 &   3.318 &   3.214 &   3.123 &   3.085 &   2.870 &   2.865 &   2.875 &   2.869 &   2.861 &   2.844 \\ 
   9500 &   2.825 &   2.819 &   2.788 &   2.788 &   2.704 &   2.702 &   2.733 &   2.734 &   2.736 &   2.731 \\ 
  10000 &   2.414 &   2.465 &   2.479 &   2.508 &   2.565 &   2.564 &   2.609 &   2.614 &   2.622 &   2.628 \\ 
  11000 &   1.748 &   1.872 &   1.945 &   2.013 &   2.331 &   2.332 &   2.396 &   2.406 &   2.422 &   2.444 \\ 
  12000 &   1.213 &   1.392 &   1.510 &   1.611 &   2.135 &   2.137 &   2.215 &   2.231 &   2.253 &   2.288 \\ 
  13000 &   0.764 &   0.988 &   1.150 &   1.283 &   1.964 &   1.967 &   2.058 &   2.079 &   2.108 &   2.154 \\ 
  15000 &   0.039 &   0.339 &   0.574 &   0.765 &   1.671 &   1.677 &   1.791 &   1.820 &   1.862 &   1.927 \\ 
  17000 &  -0.538 &  -0.183 &   0.107 &   0.342 &   1.416 &   1.425 &   1.558 &   1.595 &   1.648 &   1.729 \\ 
  19000 &  -1.025 &  -0.627 &  -0.296 &  -0.027 &   1.185 &   1.196 &   1.344 &   1.388 &   1.451 &   1.545 \\ 
  21000 &  -1.452 &  -1.016 &  -0.655 &  -0.360 &   0.972 &   0.983 &   1.144 &   1.193 &   1.262 &   1.367 \\ 
  23000 &  -1.839 &  -1.359 &  -0.974 &  -0.659 &   0.774 &   0.786 &   0.956 &   1.009 &   1.084 &   1.196 \\ 
  25000 &  -2.199 &  -1.670 &  -1.266 &  -0.934 &   0.583 &   0.594 &   0.775 &   0.831 &   0.912 &   1.032 \\ 

\\
{\boldmath ${\log(g)=4.5}$}\\
   3500 &  11.438 &  15.667 &  12.222 &  11.409 &  10.067 &   9.678 &   8.397 &   7.860 &   7.299 &   6.488 \\ 
   4000 &  10.589 &  14.012 &  10.782 &   9.907 &   8.693 &   8.320 &   7.249 &   6.775 &   6.273 &   5.722 \\ 
   4500 &   9.883 &  11.779 &   9.151 &   8.312 &   7.496 &   7.163 &   6.342 &   5.979 &   5.575 &   5.158 \\ 
   5000 &   9.214 &   9.190 &   7.549 &   6.889 &   6.569 &   6.267 &   5.625 &   5.340 &   5.013 &   4.675 \\ 
   5500 &   8.570 &   7.566 &   6.353 &   5.824 &   5.839 &   5.595 &   5.079 &   4.844 &   4.569 &   4.290 \\ 
   6000 &   7.827 &   6.369 &   5.446 &   5.016 &   5.231 &   5.048 &   4.641 &   4.446 &   4.214 &   3.982 \\ 
   6500 &   6.882 &   5.421 &   4.729 &   4.378 &   4.743 &   4.602 &   4.275 &   4.113 &   3.916 &   3.722 \\ 
   7000 &   5.920 &   4.741 &   4.271 &   4.015 &   4.279 &   4.179 &   3.930 &   3.802 &   3.646 &   3.495 \\ 
   7500 &   5.158 &   4.190 &   3.881 &   3.698 &   3.885 &   3.810 &   3.615 &   3.519 &   3.400 &   3.288 \\ 
   8000 &   4.490 &   3.738 &   3.548 &   3.426 &   3.531 &   3.478 &   3.329 &   3.263 &   3.181 &   3.105 \\ 
   8500 &   3.827 &   3.355 &   3.251 &   3.175 &   3.217 &   3.186 &   3.084 &   3.046 &   2.998 &   2.954 \\ 
   9000 &   3.207 &   3.018 &   2.975 &   2.935 &   2.966 &   2.953 &   2.893 &   2.876 &   2.854 &   2.833 \\ 
   9500 &   2.724 &   2.692 &   2.702 &   2.698 &   2.771 &   2.767 &   2.738 &   2.734 &   2.729 &   2.726 \\ 
  10000 &   2.321 &   2.388 &   2.441 &   2.468 &   2.610 &   2.612 &   2.608 &   2.613 &   2.619 &   2.629 \\ 
  11000 &   1.673 &   1.853 &   1.966 &   2.042 &   2.356 &   2.363 &   2.395 &   2.410 &   2.430 &   2.458 \\ 
  12000 &   1.154 &   1.406 &   1.560 &   1.671 &   2.154 &   2.163 &   2.218 &   2.239 &   2.268 &   2.308 \\ 
  13000 &   0.716 &   1.024 &   1.211 &   1.349 &   1.981 &   1.991 &   2.062 &   2.088 &   2.124 &   2.175 \\ 
  15000 &   0.005 &   0.404 &   0.646 &   0.833 &   1.690 &   1.703 &   1.800 &   1.834 &   1.880 &   1.949 \\ 
  17000 &  -0.560 &  -0.087 &   0.203 &   0.432 &   1.446 &   1.462 &   1.580 &   1.621 &   1.678 &   1.761 \\ 
  19000 &  -1.031 &  -0.497 &  -0.168 &   0.094 &   1.231 &   1.249 &   1.385 &   1.433 &   1.499 &   1.594 \\ 
  21000 &  -1.439 &  -0.854 &  -0.494 &  -0.204 &   1.034 &   1.054 &   1.206 &   1.259 &   1.333 &   1.438 \\ 
  23000 &  -1.804 &  -1.172 &  -0.786 &  -0.473 &   0.851 &   0.871 &   1.037 &   1.095 &   1.177 &   1.291 \\ 
  25000 &  -2.138 &  -1.461 &  -1.054 &  -0.721 &   0.677 &   0.698 &   0.876 &   0.939 &   1.027 &   1.149 \\ 
  27000 &  -2.450 &  -1.726 &  -1.300 &  -0.951 &   0.511 &   0.533 &   0.722 &   0.788 &   0.882 &   1.012 \\ 
  29000 &  -2.744 &  -1.975 &  -1.533 &  -1.167 &   0.349 &   0.372 &   0.571 &   0.642 &   0.741 &   0.878 \\ 
  31000 &  -3.015 &  -2.212 &  -1.758 &  -1.379 &   0.186 &   0.210 &   0.420 &   0.493 &   0.598 &   0.741 \\ 
  33000 &  -3.242 &  -2.425 &  -1.964 &  -1.578 &   0.031 &   0.055 &   0.273 &   0.349 &   0.457 &   0.605 \\ 
  35000 &  -3.418 &  -2.597 &  -2.134 &  -1.744 &  -0.101 &  -0.077 &   0.147 &   0.223 &   0.331 &   0.481 \\ 
  40000 &  -3.716 &  -2.868 &  -2.401 &  -2.009 &  -0.319 &  -0.296 &  -0.064 &   0.011 &   0.116 &   0.267 \\ 
\enddata
\end{deluxetable}

\begin{deluxetable}{ccccccccccc}
\tablecolumns{11}
\tablecaption{Extinction coefficients $\mathcal{R}$ in selected \wfpc\ broad
band filters from the \protect{\citet{bes98}} model atmospheres and the
Galactic reddening law of \protect{\citet{savmat79}} as a function of effective
temperature and \ebv (see equation~(\ref{eq:redmag})).\label{tab:abs}}

\tablehead{\ebv & \te=3500~K & 4000 & 5250 & 6500 & 8750 & 9250 & 10000 & 25000 & 40000}
\startdata

\\
{\bf F170W}\\
 0.0 &  2.017 &  2.205 &  3.060 &  6.202 &  7.901 &  7.956 &  8.009 &  8.110 &  8.111 \\ 
 0.2 &  1.970 &  2.143 &  2.851 &  5.484 &  7.717 &  7.818 &  7.908 &  8.080 &  8.086 \\ 
 0.4 &  1.927 &  2.088 &  2.689 &  4.812 &  7.427 &  7.598 &  7.746 &  8.045 &  8.058 \\ 
 0.6 &  1.889 &  2.039 &  2.559 &  4.265 &  6.993 &  7.248 &  7.480 &  7.998 &  8.023 \\ 
 0.8 &  1.854 &  1.994 &  2.452 &  3.849 &  6.438 &  6.757 &  7.073 &  7.926 &  7.973 \\ 
 1.0 &  1.822 &  1.954 &  2.361 &  3.534 &  5.856 &  6.192 &  6.550 &  7.799 &  7.892 \\ 
 1.2 &  1.793 &  1.916 &  2.284 &  3.289 &  5.331 &  5.649 &  6.003 &  7.578 &  7.745 \\ 
 1.4 &  1.766 &  1.882 &  2.216 &  3.095 &  4.893 &  5.181 &  5.508 &  7.232 &  7.491 \\ 
 1.5 &  1.753 &  1.866 &  2.185 &  3.012 &  4.704 &  4.977 &  5.288 &  7.021 &  7.319 \\ 

{\bf F255W}\\
 0.0 &  4.699 &  5.262 &  6.256 &  6.548 &  6.772 &  6.791 &  6.812 &  6.934 &  6.958 \\ 
 0.2 &  4.437 &  5.095 &  6.214 &  6.494 &  6.713 &  6.731 &  6.751 &  6.871 &  6.895 \\ 
 0.4 &  4.147 &  4.892 &  6.171 &  6.443 &  6.657 &  6.675 &  6.696 &  6.813 &  6.836 \\ 
 0.6 &  3.848 &  4.652 &  6.125 &  6.396 &  6.606 &  6.623 &  6.644 &  6.759 &  6.781 \\ 
 0.8 &  3.562 &  4.385 &  6.072 &  6.350 &  6.557 &  6.575 &  6.595 &  6.709 &  6.731 \\ 
 1.0 &  3.304 &  4.108 &  6.005 &  6.303 &  6.510 &  6.528 &  6.549 &  6.662 &  6.684 \\ 
 1.2 &  3.081 &  3.840 &  5.914 &  6.252 &  6.463 &  6.482 &  6.503 &  6.619 &  6.639 \\ 
 1.4 &  2.892 &  3.595 &  5.787 &  6.192 &  6.414 &  6.434 &  6.457 &  6.577 &  6.598 \\ 
 1.5 &  2.809 &  3.483 &  5.707 &  6.156 &  6.388 &  6.408 &  6.433 &  6.557 &  6.578 \\ 

{\bf F300W}\\
 0.0 &  3.789 &  4.339 &  5.173 &  5.391 &  5.603 &  5.628 &  5.660 &  5.851 &  5.883 \\ 
 0.2 &  3.577 &  4.185 &  5.131 &  5.353 &  5.558 &  5.582 &  5.614 &  5.804 &  5.836 \\ 
 0.4 &  3.360 &  4.004 &  5.079 &  5.314 &  5.513 &  5.537 &  5.569 &  5.759 &  5.791 \\ 
 0.6 &  3.152 &  3.804 &  5.014 &  5.271 &  5.467 &  5.492 &  5.524 &  5.716 &  5.747 \\ 
 0.8 &  2.965 &  3.598 &  4.931 &  5.221 &  5.419 &  5.444 &  5.477 &  5.674 &  5.705 \\ 
 1.0 &  2.802 &  3.399 &  4.822 &  5.160 &  5.365 &  5.391 &  5.427 &  5.632 &  5.665 \\ 
 1.2 &  2.664 &  3.217 &  4.685 &  5.083 &  5.301 &  5.330 &  5.370 &  5.591 &  5.624 \\ 
 1.4 &  2.547 &  3.055 &  4.520 &  4.982 &  5.222 &  5.255 &  5.301 &  5.549 &  5.584 \\ 
 1.5 &  2.496 &  2.982 &  4.429 &  4.921 &  5.175 &  5.211 &  5.260 &  5.526 &  5.563 \\ 

{\bf F336W}\\
 0.0 &  4.045 &  4.412 &  4.911 &  4.974 &  4.998 &  5.002 &  5.007 &  5.037 &  5.041 \\ 
 0.2 &  3.888 &  4.296 &  4.888 &  4.962 &  4.990 &  4.994 &  5.000 &  5.033 &  5.037 \\ 
 0.4 &  3.721 &  4.160 &  4.857 &  4.948 &  4.979 &  4.984 &  4.991 &  5.028 &  5.033 \\ 
 0.6 &  3.554 &  4.008 &  4.816 &  4.929 &  4.965 &  4.971 &  4.979 &  5.022 &  5.028 \\ 
 0.8 &  3.397 &  3.848 &  4.761 &  4.903 &  4.947 &  4.954 &  4.964 &  5.016 &  5.022 \\ 
 1.0 &  3.254 &  3.689 &  4.690 &  4.868 &  4.922 &  4.931 &  4.944 &  5.009 &  5.016 \\ 
 1.2 &  3.130 &  3.539 &  4.599 &  4.821 &  4.888 &  4.900 &  4.917 &  5.000 &  5.009 \\ 
 1.4 &  3.023 &  3.403 &  4.489 &  4.759 &  4.842 &  4.858 &  4.880 &  4.988 &  4.999 \\ 
 1.5 &  2.976 &  3.341 &  4.428 &  4.721 &  4.814 &  4.831 &  4.857 &  4.980 &  4.994 \\ 

{\bf F439W}\\
 0.0 &  4.088 &  4.097 &  4.132 &  4.152 &  4.161 &  4.163 &  4.166 &  4.178 &  4.181 \\ 
 0.2 &  4.085 &  4.094 &  4.129 &  4.149 &  4.159 &  4.161 &  4.164 &  4.176 &  4.179 \\ 
 0.4 &  4.083 &  4.092 &  4.127 &  4.147 &  4.156 &  4.158 &  4.161 &  4.173 &  4.177 \\ 
 0.6 &  4.080 &  4.089 &  4.124 &  4.144 &  4.154 &  4.156 &  4.159 &  4.171 &  4.174 \\ 
 0.8 &  4.076 &  4.086 &  4.121 &  4.142 &  4.152 &  4.154 &  4.156 &  4.168 &  4.172 \\ 
 1.0 &  4.073 &  4.083 &  4.118 &  4.139 &  4.149 &  4.151 &  4.154 &  4.166 &  4.169 \\ 
 1.2 &  4.069 &  4.079 &  4.116 &  4.136 &  4.147 &  4.149 &  4.151 &  4.163 &  4.167 \\ 
 1.4 &  4.065 &  4.076 &  4.113 &  4.134 &  4.144 &  4.146 &  4.149 &  4.161 &  4.164 \\ 
 1.5 &  4.062 &  4.074 &  4.111 &  4.132 &  4.143 &  4.145 &  4.148 &  4.160 &  4.163 \\ 

{\bf F450W}\\
 0.0 &  3.787 &  3.798 &  3.835 &  3.890 &  3.929 &  3.935 &  3.944 &  3.987 &  4.000 \\ 
 0.2 &  3.779 &  3.790 &  3.825 &  3.880 &  3.918 &  3.925 &  3.933 &  3.976 &  3.989 \\ 
 0.4 &  3.772 &  3.782 &  3.816 &  3.869 &  3.908 &  3.914 &  3.922 &  3.965 &  3.977 \\ 
 0.6 &  3.764 &  3.774 &  3.806 &  3.859 &  3.897 &  3.903 &  3.912 &  3.953 &  3.966 \\ 
 0.8 &  3.757 &  3.767 &  3.797 &  3.848 &  3.887 &  3.893 &  3.901 &  3.942 &  3.954 \\ 
 1.0 &  3.749 &  3.759 &  3.789 &  3.838 &  3.876 &  3.882 &  3.890 &  3.930 &  3.942 \\ 
 1.2 &  3.742 &  3.752 &  3.780 &  3.828 &  3.866 &  3.872 &  3.880 &  3.919 &  3.931 \\ 
 1.4 &  3.735 &  3.745 &  3.771 &  3.818 &  3.856 &  3.862 &  3.869 &  3.908 &  3.919 \\ 
 1.5 &  3.731 &  3.741 &  3.767 &  3.814 &  3.851 &  3.857 &  3.864 &  3.902 &  3.914 \\ 

{\bf F555W}\\
 0.0 &  3.031 &  3.066 &  3.153 &  3.189 &  3.243 &  3.249 &  3.255 &  3.281 &  3.288 \\ 
 0.2 &  3.019 &  3.053 &  3.139 &  3.175 &  3.228 &  3.235 &  3.241 &  3.267 &  3.274 \\ 
 0.4 &  3.008 &  3.041 &  3.126 &  3.162 &  3.214 &  3.220 &  3.227 &  3.252 &  3.259 \\ 
 0.6 &  2.997 &  3.030 &  3.113 &  3.148 &  3.200 &  3.206 &  3.212 &  3.238 &  3.245 \\ 
 0.8 &  2.987 &  3.018 &  3.100 &  3.135 &  3.186 &  3.193 &  3.199 &  3.224 &  3.230 \\ 
 1.0 &  2.977 &  3.007 &  3.088 &  3.122 &  3.173 &  3.179 &  3.185 &  3.210 &  3.216 \\ 
 1.2 &  2.967 &  2.997 &  3.076 &  3.110 &  3.159 &  3.166 &  3.172 &  3.196 &  3.203 \\ 
 1.4 &  2.958 &  2.987 &  3.064 &  3.097 &  3.146 &  3.153 &  3.158 &  3.183 &  3.189 \\ 
 1.5 &  2.953 &  2.982 &  3.058 &  3.091 &  3.140 &  3.146 &  3.152 &  3.176 &  3.182 \\ 

{\bf F606W}\\
 0.0 &  2.712 &  2.729 &  2.805 &  2.841 &  2.893 &  2.900 &  2.906 &  2.932 &  2.938 \\ 
 0.2 &  2.703 &  2.719 &  2.793 &  2.828 &  2.879 &  2.886 &  2.892 &  2.918 &  2.924 \\ 
 0.4 &  2.694 &  2.709 &  2.781 &  2.815 &  2.866 &  2.873 &  2.879 &  2.904 &  2.910 \\ 
 0.6 &  2.685 &  2.700 &  2.770 &  2.803 &  2.853 &  2.860 &  2.866 &  2.891 &  2.897 \\ 
 0.8 &  2.677 &  2.691 &  2.759 &  2.791 &  2.841 &  2.847 &  2.853 &  2.877 &  2.883 \\ 
 1.0 &  2.669 &  2.682 &  2.748 &  2.780 &  2.828 &  2.834 &  2.840 &  2.864 &  2.870 \\ 
 1.2 &  2.661 &  2.674 &  2.737 &  2.768 &  2.816 &  2.822 &  2.828 &  2.852 &  2.857 \\ 
 1.4 &  2.654 &  2.665 &  2.727 &  2.758 &  2.805 &  2.810 &  2.816 &  2.839 &  2.844 \\ 
 1.5 &  2.650 &  2.661 &  2.722 &  2.752 &  2.799 &  2.805 &  2.810 &  2.833 &  2.838 \\ 

{\bf F675W}\\
 0.0 &  2.430 &  2.432 &  2.444 &  2.450 &  2.460 &  2.461 &  2.462 &  2.468 &  2.470 \\ 
 0.2 &  2.427 &  2.429 &  2.441 &  2.447 &  2.457 &  2.458 &  2.460 &  2.465 &  2.467 \\ 
 0.4 &  2.425 &  2.426 &  2.439 &  2.445 &  2.454 &  2.456 &  2.457 &  2.463 &  2.464 \\ 
 0.6 &  2.422 &  2.424 &  2.436 &  2.442 &  2.452 &  2.453 &  2.454 &  2.460 &  2.461 \\ 
 0.8 &  2.419 &  2.421 &  2.433 &  2.439 &  2.449 &  2.450 &  2.451 &  2.457 &  2.459 \\ 
 1.0 &  2.416 &  2.418 &  2.430 &  2.436 &  2.446 &  2.447 &  2.449 &  2.455 &  2.456 \\ 
 1.2 &  2.413 &  2.416 &  2.428 &  2.434 &  2.443 &  2.445 &  2.446 &  2.452 &  2.453 \\ 
 1.4 &  2.411 &  2.413 &  2.425 &  2.431 &  2.441 &  2.442 &  2.443 &  2.449 &  2.451 \\ 
 1.5 &  2.409 &  2.412 &  2.424 &  2.430 &  2.439 &  2.441 &  2.442 &  2.448 &  2.449 \\ 

{\bf F814W}\\
 0.0 &  1.786 &  1.829 &  1.860 &  1.875 &  1.892 &  1.894 &  1.896 &  1.912 &  1.916 \\ 
 0.2 &  1.777 &  1.820 &  1.851 &  1.867 &  1.884 &  1.886 &  1.888 &  1.904 &  1.908 \\ 
 0.4 &  1.768 &  1.811 &  1.842 &  1.858 &  1.876 &  1.877 &  1.879 &  1.896 &  1.900 \\ 
 0.6 &  1.759 &  1.801 &  1.833 &  1.849 &  1.867 &  1.868 &  1.870 &  1.888 &  1.892 \\ 
 0.8 &  1.750 &  1.792 &  1.824 &  1.840 &  1.858 &  1.859 &  1.861 &  1.879 &  1.884 \\ 
 1.0 &  1.740 &  1.782 &  1.815 &  1.831 &  1.849 &  1.850 &  1.852 &  1.870 &  1.875 \\ 
 1.2 &  1.731 &  1.773 &  1.805 &  1.822 &  1.839 &  1.841 &  1.843 &  1.861 &  1.866 \\ 
 1.4 &  1.721 &  1.763 &  1.795 &  1.812 &  1.830 &  1.832 &  1.833 &  1.852 &  1.857 \\ 
 1.5 &  1.717 &  1.758 &  1.791 &  1.807 &  1.825 &  1.827 &  1.829 &  1.848 &  1.853 \\ 
\enddata
\end{deluxetable}

\end{document}